\documentclass[nofootinbib]{revtex4-1}
\usepackage[utf8]{inputenc}
\usepackage{amsmath}
\usepackage{graphicx}
\usepackage{amssymb}
\usepackage{color,colortbl}
\usepackage{cancel}
\usepackage{framed}
\usepackage{comment}
\usepackage[margin=0.7in]{geometry}
\usepackage{mathtools}
\usepackage{hyperref}
\usepackage{chngcntr}
\usepackage{soul}
\usepackage{subfig}
\usepackage{marvosym}

\counterwithin{equation}{section}

\makeatletter

    \def\CT@@do@color{%
      \global\let\CT@do@color\relax
            \@tempdima\wd\z@
            \advance\@tempdima\@tempdimb
            \advance\@tempdima\@tempdimc
    \advance\@tempdimb\tabcolsep
    \advance\@tempdimc\tabcolsep
    \advance\@tempdima2\tabcolsep
            \kern-\@tempdimb
            \leaders\vrule
                    \hskip\@tempdima\@plus  1fill
            \kern-\@tempdimc
            \hskip-\wd\z@ \@plus -1fill }
    \makeatother

\newcommand{\beq}{\begin{equation}}
\newcommand{\eeq}{\end{equation}}
\newcommand{\bea}{\begin{eqnarray}}
\newcommand{\eea}{\end{eqnarray}}

\begin{document}


\date{\today}

\title{\Large \bfseries Using atomic clocks to detect local dark matter halos}
\author{Chris Kouvaris}
\email{kouvaris@mail.ntua.gr}
\affiliation{National Technical University of Athens, Zografou Campus 9, Heroon Polytechneiou Str. 157 80 Athens, Greece}
\author{Eleftherios Papantonopoulos}
\email{lpapa@central.ntua.gr}
\affiliation{National Technical University of Athens, Zografou Campus 9, Heroon Polytechneiou Str. 157 80 Athens, Greece}
\author{Lauren Street}
\email{streetlg@mail.uc.edu}
\affiliation{University of Cincinnati, Dept. of Physics, Cincinnati, OH 45221 USA}
\author{L.C.R. Wijewardhana}
\email{rohana.wijewardhana@gmail.com}
\affiliation{University of Cincinnati, Dept. of Physics, Cincinnati, OH 45221 USA}

\begin{abstract}
It is possible that bosonic dark matter forms halos around the Sun or the Earth.  We discuss the possibility of probing such halos with atomic clocks.  Focusing on either a Higgs portal or photon portal interaction between the dark matter and the Standard Model, we search the possible parameter space for which a clock on Earth and a clock in space would have a discernible frequency difference.  Bosonic dark matter halos surrounding the Earth can potentially be probed with current optical atomic clocks.
\end{abstract}

\maketitle

\section{Introduction}
Although there is solid evidence for the existence of dark matter (DM) through its gravitational effects on the universe (e.g. the cosmic Microwave Background~\cite{Aghanim:2018eyx}), there is very little information regarding the nature of DM other than its contribution to matter and consequently  gravity. DM candidates range from primordial black holes \cite{Frampton_2009,Frampton_2010,Garc_a_Bellido_2017} to axions or axion like particles (ALPs) with masses as small as $10^{-22}\text{eV}$ \cite{Peccei:1977hh,Peccei:1977ur,Weinberg:1977ma,Wilczek:1977pj,DINE1981199,Kim:1979if,SHIFMAN1980493,Turner:1983he,Press:1989id,Sin:1992bg,Hu:2000ke,GOODMAN2000103,Peebles_2000,AMENDOLA2006192,Arvanitaki:2009fg,Li:2013nal,MARSH20161,Hui:2016ltb,Lee:2017qve}. For many years the most popular and well motivated DM paradigm was that of  the Weakly Interacting Massive Particles (WIMPs), where the observed relic abundance of DM is adjusted by DM annihilations that take place in the early universe. An annihilation cross section on the order of the weak interactions provides the observed DM abundance. WIMP models, however, are constrained in several ways, for example from expected direct scattering between WIMPs and Earth based detectors. Although the DM annihilation  cross section can be quite different from the DM-nucleon or DM-electron cross section which is relevant in direct detection, current sophisticated direct search experiments have set unprecedented limits which have made the WIMP paradigm less appealing (see e.g. \cite{Aprile:2018dbl}). Furthermore, there has been a systematic study of indirect detection channels of DM discovery. Although the products of DM annihilation vary accordingly to the underlying WIMP model, there is an extensive and thorough investigation of a vast class of WIMP models. These dedicated searches focus, for example, on places where DM could be present in large quantities such as the center of our galaxy and  set constraints on different annihilation channels of DM, especially to hadrons and photons. The inconclusive/negative findings further limit the parameter space where the WIMP paradigm can account for the relic abundance of DM without fine tuning of the model parameters (see \cite{Slatyer:2017sev} for an overview of the subject). Therefore, alternative classes of DM models have gained interest including, among others, asymmetric DM.  In this DM paradigm, a mechanism, similar to the baryon case, creates an asymmetry between the number of particles and antiparticles with subsequent annihilations depleting one of the two species, leaving only one of the species in excess \cite{Petraki:2013wwa,Pollack_2015,Chang_2019,Kouvaris:2015rea,Alonso-Alvarez:2019pfe,Maselli:2019ubs}. In such a mechanism, after the depletion of the minority species, no substantial annihilations take place and therefore no signal is expected from DM annihilation at the center of galaxy or elsewhere. In general, sufficiently light DM particles can easily evade direct detection because even with a  large DM-nucleon/electron cross section, the produced recoil energy falls below the energy threshold that triggers the detector.     

Given the current lack of evidence of WIMPs, lighter particles and alternative mechanisms of DM production should be looked for. One example that falls in this category is the axion, a well motivated particle that potentially solves the strong CP problem of QCD. Axions and ALPs are too light to be detected using direct detection methods based on recoil and therefore alternative detection methods have been implemented. New, promising, and relatively unexplored tools that can facilitate the exploration of new physics are atomic clocks. Providing unprecedented accuracy, atomic clocks can be used in different contexts to probe new physics and/or particular DM candidates. The ticking rate of atomic clocks is governed, in general, by fundamental constants such as the fine structure constant $\alpha$, and the proton and electron masses. Even tiny changes to these constants, due to new physics, could create a detectable change in the ticking rate of the clock. Since there is no absolute clock in nature, the potential detection of new physics is related to comparison in the ticking rate among different clocks.  These clocks are usually separated by some distance in order to allow new physics effects to influence the separated clocks in different ways. In this paper we are interested in potential applications for DM detection. One way that atomic clocks can be affected is through extensive objects that change vacuum properties of the environment surrounding the clocks.  This, in turn, causes changes in the aforementioned constants and  creates de-sychronizations of clocks as the objects pass through the Earth engulfing different clocks at different times. In particular, using data collected over several years from GPS satellite based atomic clocks,  the authors of \cite{Derevianko:2013oaa,Roberts:2017hla,Afach:2021pfd} managed to set constraints on  models where DM is in the form of topological defects that occasionally  pass through the Earth. These topological defects alter 
$\alpha$ and, therefore, as the topological defect passes through the Earth, there should be  a geometric pattern in the  change of each clock relative to a reference clock. There has, since this study, been several others that exploit atomic clocks to probe DM either by focusing on the transient effect of bypassing DM structures or by effects caused by the time variation of the DM field~\cite{Arvanitaki:2014faa,Stadnik:2014tta,VanTilburg:2015oza,Hees:2016gop,Hees:2018fpg,Roberts:2018agv,Roberts:2018xqn,Wolf:2018xlz,Alonso:2018dxy,Savalle:2019jsb}.

It is possible that QCD axions and ALPs form bound Bose-Einstein condensate structures called boson stars, while extensive studies have been performed to understand these structures \cite{Kaup:1968zz,Ruffini:1969qy,BREIT1984329,Colpi:1986ye,Seidel:1990jh,Friedberg:1986tq,Seidel:1991zh,LEE1992251,Guzman:2006yc,PhysRevD.68.023511,PhysRevD.53.2236,Matos:2007zza,Bernal:2009zy,UrenaLopez:2010ur,Chavanis:2011zi,Chavanis:2011zm,doi:10.1142/S0217732316500905,Eby:2016cnq,Eby:2017azn,VISINELLI201864,Levkov:2018kau,Lin:2018whl,Guzman:2019gqc,Braaten:2019knj,sym12010025,Eby:2019ntd,Eggemeier:2019jsu,Kirkpatrick:2020fwd,Eby:2020ply}.  In a previous paper \cite{Kouvaris:2019nzd}, we explored the possibility of detecting dilute asymmetric bosonic DM stars that pass through the Earth.  In this case, we assumed the boson stars were composed of light bosons that couple to the Standard Model particles via a Higgs or a photon portal. As the star engulfs a clock, the Higgs (photon) portal induces a change in the electron mass (fine structure constant) which leads to a different clock metronomy. In this paper we are not interested in transient effects of dilute bosonic stars. Instead, we focus on the possibility that the Sun and/or the Earth have respective halos composed of light bosons that again exhibit couplings to the SM particles via a Higgs or photon portal. A similar setup has been studied in the case of relaxions in~\cite{Banerjee:2019epw}. In Sec. \ref{sec:bsts} we present the profile of the bosonic halo that surrounds the Earth and the Sun. In Sec. \ref{sec:portals} we present the two portals and how the portals affect the metronomy of the atomic clocks in Sec. \ref{sec:freqshifts}. Finally, the results and conclusions are presented in Sec. \ref{sec:results}.

\section{Boson star formalism}\label{sec:bsts}
We consider the possibility that a dilute boson star surrounds an external gravitational source, namely the Sun or the Earth.  Dilute boson stars are well known to be both structurally stable and stable to decay \cite{Chavanis:2011zi,Chavanis:2011zm,doi:10.1142/S0217732316500905,Eby:2017azn}.  The energy functional of a dilute boson star subject to an external gravitational source is given by,
\begin{widetext}
\begin{align} \label{eq:energy_func}
E[\psi] = \int d^3 r \left[\frac{\left|\nabla \psi\right|^2}{2m} + \frac{m}{2} \Phi_g\left|\psi\right|^2 - \frac{|\lambda|}{16 m^2}|\psi|^4 + m \, \Phi_{g,\text{ext}} |\psi|^2 \right],
\end{align}
\end{widetext}
where $\Phi_g$ is the Newtonian self-gravitational potential satisfying the Poisson equation,
\begin{align}
\nabla^2 \Phi_g = 4 \pi \frac{m}{M_P^2} \left|\psi\right|^2,
\end{align}
with $M_P \approx 1.22\times 10^{19} \, \text{GeV}$ the Planck mass and $\lambda$ the self-interaction coupling of the bosons where we take the self-interactions to be attractive.  In this case, we take the $|\psi|^4$ term to be the dominant term in the self-interaction potential of the scalar field.  However, we note that higher order terms bound the potential and gives rise to the possibility of ``dense" configurations \cite{Kolb:1993hw,doi:10.1142/S0217732316500905,Braaten:2019knj,VISINELLI201864}.  The energy term due to the external gravitational source is \cite{Eby:2020eas},
\begin{align}
E_{g,\text{ext}} = m \int d^3 r\, \Phi_{g,\text{ext}} |\psi|^2,
\end{align}
where $m$ is the boson mass, $\psi$ is the wavefunction of the boson star, and the gravitational potential due to the external source is taken to be,
\begin{align}
\Phi_{g,\text{ext}} = -\frac{M_*}{M_P^2}
	\times 
\begin{cases}
\displaystyle{\frac{3}{2 R_*} - \frac{r^2}{2R_*^3}} & \text{for}\quad r \leq R_*
\\ \\
\displaystyle{\frac{1}{r}} & \text{for} \quad r > R_*
\end{cases},
\end{align}
where $M_*$ is the mass and $R_*$ the radius of the external source.

We take the wavefunction to be an ansatz of the form,
\begin{align}\label{eq:linexp}
\psi(r) = \sqrt{\frac{N}{7 \pi \sigma^3}}\left(1 + \frac{r}{\sigma}\right) \exp\left(-\frac{r}{\sigma}\right),
\end{align}
which was found to be a good approximation to numerical solutions for dilute boson stars not subject to an external gravitational potential in \cite{Eby:2018dat}.  Here, $\sigma$ is some variational parameter to be found by minimzing the energy functional (Eq. \ref{eq:energy_func}) and $N$ is the total particle number of the boson star. For this ansatz, the Newtonian self-gravitational potential is,
\begin{widetext}
\begin{align}
\Phi_g = -\frac{m N}{M_P^2} \frac{1}{14 \, r}\left\{14 - \exp\left(- \frac{2 r}{\sigma}\right)\left[2+\frac{r}{\sigma}\right]\left[2\left(\frac{r}{\sigma}\right)^2 + 6\left(\frac{r}{\sigma}\right) + 7\right]\right\}.
\end{align}
\end{widetext}

We scale the radius and mass of the boson star, as well as those of the external source in order to obtain dimensionless parameters as,
\begin{align}\label{eq:scaling}
\begin{cases}
\sigma = \sqrt{|\lambda|}\frac{M_P}{m^2}\tilde{r},
&
M = \frac{M_P}{\sqrt{|\lambda|}}n,
\\
R_* = \sqrt{|\lambda|}\frac{M_P}{m^2}\tilde{r}_*,
&
M_* = \frac{M_P}{\sqrt{|\lambda|}}n_*.
\end{cases}
\end{align}
Using the ansatz of Eq. (\ref{eq:linexp}), the total energy given by Eq. (\ref{eq:energy_func}) per particle number now evaluates to,
\begin{align}\label{eq:Etot}
\frac{E}{N} = \frac{m^2}{M_P \sqrt{|\lambda|^3}} \left(\frac{a}{\tilde{r}^2} - \frac{b \, n}{\tilde{r}}- \frac{c \, n}{\tilde{r}^3}\right)  + \frac{E_{g,\text{ext}}}{N},
\end{align}
where $\tilde{r}$ and $n$ are the scaled radius and particle number of the boson star given by Eq. (\ref{eq:scaling}) and $a$, $b$, $c$ are constants given by,
\begin{align}
a = \frac{3}{14}, \qquad b=\frac{5373}{25088}, \qquad c=\frac{437}{200704\pi}.
\end{align}
The first term in the parentheses of Eq. (\ref{eq:Etot}) is the kinetic energy, the second is the self-gravitational energy, and the third is the self-interaction energy.  The total gravitational energy per particle number due to the external source is given by,
\begin{align}\label{eq:egtot}
\frac{E_{g,\text{ext}}}{N} = \frac{3 \, n_* \, \tilde{r}^2}{28 \, \tilde{r}_*^3} \left\{27 - 14 \left(\frac{\tilde{r}_*}{\tilde{r}}\right)^2 - \exp\left(-\frac{2 \tilde{r}_*}{\tilde{r}}\right) \left[ 27 + 54 \left(\frac{\tilde{r}_*}{\tilde{r}}\right) + 40 \left(\frac{\tilde{r}_*}{\tilde{r}}\right)^2 + 14 \left(\frac{\tilde{r}_*}{\tilde{r}}\right)^3 + 2 \left(\frac{\tilde{r}_*}{\tilde{r}}\right)^4 \right] \right\},
\end{align}
where $\tilde{r}_*$ and $n_*$ are the scaled radius and mass of the external source given by Eq. (\ref{eq:scaling}).

Now, the scaled variational parameter $\tilde{r}$ can be found by minimizing the total energy (Eqs. (\ref{eq:Etot}) and (\ref{eq:egtot})), and can be found exactly through numerical methods.  However, we choose to keep the minimization process analytic by expanding the total external gravitational energy (Eq. (\ref{eq:egtot})) to a particular order of $\tilde{r}_*/\tilde{r}$.  The order of expansion taken is such that the fractional energy difference from the total external gravitational energy is a small number (for example $10^{-3}$),
\begin{align}\label{eq:egext_diff}
\frac{E_{g,\text{ext}} - E_{g,\text{ext}}^\text{approx}}{E_{g,\text{ext}}} \lesssim 10^{-3},
\end{align}
where $E_{g,\text{ext}}^\text{approx}$ is the total external gravitational energy of Eq. (\ref{eq:egtot}) expanded to some power of $\tilde{r}_*/\tilde{r}$.  We take the possible scaled radius of the bosonic dark matter halo to be $\tilde{r} \geq \tilde{r}_*$.  However, in order the ensure the expansion condition (Eq. (\ref{eq:egext_diff})), we must split the possible scaled radius into two regions, where we expand the total external gravitational energy (Eq. (\ref{eq:egtot})) to different orders of $\tilde{r}_*/\tilde{r}$.  For the case $\tilde{r} \geq 2 \tilde{r}_*$, we need to expand to $\mathcal{O}\left[\left(\tilde{r}_*/\tilde{r}\right)^2\right]$ in order to ensure Eq. (\ref{eq:egext_diff}).  In this case, the approximate gravitational energy due to the external source is,
\begin{align}\label{eq:egapp1}
\left(\frac{E_{g,\text{ext}}}{N}\right)_{I} \approx \frac{n_*}{\tilde{r}}\left[-b'+ c' \left(\frac{\tilde{r}_*}{\tilde{r}}\right)^2\right],
\end{align}
where $b' = 9/14$ and $c' = 2/35$.  One can see that the first and second term of this approximate external gravitational energy contributes to the total energy as the self-gravitational energy and self-interaction energy of Eq. (\ref{eq:Etot}), respectively.   For $\tilde{r}_* \leq \tilde{r} < 2 \tilde{r}_*$ we need to expand to $\mathcal{O}\left[\left(\tilde{r}_*/\tilde{r}\right)^5\right]$  in order to ensure Eq. (\ref{eq:egext_diff}) which gives the approximate gravitational energy due to the external source,
\begin{align}\label{eq:egapp2}
\left(\frac{E_{g,\text{ext}}}{N}\right)_{II} \approx \left(\frac{E_{g,\text{ext}}}{N}\right)_{I} + \frac{n_*}{\tilde{r}}\left[-\frac{3}{245} \left(\frac{\tilde{r}_*}{\tilde{r}}\right)^4 + \frac{1}{210} \left(\frac{\tilde{r}_*}{\tilde{r}}\right)^5 \right].
\end{align}

For the approximate external gravitational energy of Eq. (\ref{eq:egapp1}), the solution $\left(\tilde{r}_d\right)_{I}$ for which the total energy (Eq. (\ref{eq:Etot})) is minimized is,
\begin{align}\label{eq:rhosol}
\left(\tilde{r}_d\right)_{I} = \frac{a}{b n + b' n_*}\left[1 + \sqrt{1 - \frac{3\left(b n +b' n_*\right)(c n - c' n_* \tilde{r}_*^2)}{a^2}}\right].
\end{align}
There is a critical particle number beyond which no stable states exist,
\begin{align}\label{eq:ncrit}
\left(n_c\right)_{I} = \sqrt{\frac{a^2 + 3 b' c' n_*^2 \tilde{r}_*^2}{3 b c}} \left\{\sqrt{\frac{3}{4 b c}}\frac{b c' n_* \tilde{r}_*^2 - b' c n_*}{\sqrt{a^2 + 3 b' c' n_*^2 \tilde{r}_*^2}} + \sqrt{1 + \frac{3}{4 b c}\frac{(b c' n_* \tilde{r}_*^2 - b' c n_*)^2}{a^2 + 3 b' c' n_*^2 \tilde{r}_*^2}}\right\}.
\end{align}
It is instructive to compare the properties of a boson star subject to an external gravitational source to those of a boson star not subject to any external gravity.  One can see that for $n_* = 0$ and $\tilde{r}_* = 0$ (i.e. removing the external source) \cite{Chavanis:2011zi,Eby:2020eas}, the minimum energy solution and critical particle number reduce to their usual forms for a boson star not subject to an external gravitational source,
\begin{align}
\left(\tilde{r}_d\right)_{I}\Big|_{n_*,\tilde{r}_*=0} = \frac{a}{b n} \left[1 + \sqrt{1 - \left(\frac{n}{n_c}\right)^2}\right],
\qquad
\left(n_c\right)_{I} \Big|_{n_*,\tilde{r}_*=0} =  \frac{a}{\sqrt{3 b c}}.
\end{align}
For the approximate external gravitational energy of Eq. (\ref{eq:egapp2}), the total energy can also be minimized to find a stable minimum energy solution which we define as $\left(\tilde{r}_d\right)_{II}$.  

Finally, the density of the boson star at some scaled radius $x$ from the center of the boson star is,
\begin{align}\label{eq:vphi}
\rho_\phi(x) = m \, \psi^2(x).
\end{align}
This is the parameter which dictates the frequency shift of an atomic clock as it sits at some position $x$ inside the boson star.  Also, we take the radius of the boson star to be equal to $R_{99}$, the radius inside which $99\%$ of the mass is contained.  For the ansatz chosen, $R_{99}$ is related to the minimum energy solution as,
\begin{align}
R_{99} \approx 5 \sigma_d \approx 5 \sqrt{|\lambda|}\frac{M_P}{m^2} \tilde{r}_d
\end{align}

\subsection{Constraints}\label{bst_dens_consts}
In this analysis, we consider the possibility of both solar bound and Earth bound boson stars.  For both possibilities, the parameter space can be constrained from experimental evidence of the gravitational influence of dark matter in our solar system or near the Earth.  For a solar bound halo, an upper limit has been placed on the density of DM within our solar system from the EPM2011 solar system ephemerides \cite{Pitjev:2013sfa}.  These constraints give the maximum density that can be within a given radius from the center of the Sun.  The most constraining of these are shown in Table \ref{tab:bst_dens_consts} where $\rho_\text{DM,max}$ is the upper limit on the total DM density at a given radius.
\begin{table}[h!]
  \begin{center}
    \caption{Constraints on the total dark matter density in the solar system at a given radius}
    \label{tab:bst_dens_consts}
    \begin{tabular}{|c||c|c|c|c|c|}
	\hline
	Planet & Mercury & Venus & Earth & Mars & Saturn
	\\ \hline
      Radius $[\text{AU}]$ & $0.387098$ 
		& $0.723332$
		& $1.000000$ 
		& $1.523660$ 
		& $9.582420$ 
	\\ \hline
	$\rho_\text{DM,max} \, [\text{g} \, \text{cm}^{-3}]$
		& $9.3 \times 10^{-18}$ 
		& $1.9 \times 10^{-18}$
		& $1.4 \times 10^{-19}$
		& $1.4 \times 10^{-20}$
		& $1.1 \times 10^{-20}$
	\\ \hline	
    \end{tabular}
  \end{center}
\end{table}

For Earth bound boson stars, an upper limit has been placed on the total mass that can be present between the orbital radius of the moon ($\sim 384,000\,\text{km}$) and the Laser Geometric Environmental Observation Survey (LAGEOS) satellites ($\sim 12,300 \, \text{km}$) of \cite{Adler:2008rq}
\begin{align}\label{eq:massconst_earth}
M_\text{enc} < 4 \times 10^{-9} \, M_\oplus
\end{align}.

\section{Portals to Standard Model}\label{sec:portals}
\subsection{Higgs portal}\label{sec:higgs_portal}
We consider the interaction between the DM, quanta of a complex scalar field $\phi$, and the SM through a Higgs portal \cite{Piazza:2010ye,Stadnik:2016zkf,Flacke:2016szy,Cosme:2018nly,Alonso-Alvarez:2019pfe}.
\begin{align}
\mathcal{L} = ... + \beta \left|\phi\right|^2 \left|H\right|^2,
\end{align}
where $\beta$, a positive constant, is the coupling constant between the Higgs and the DM.  This interaction induces a shift in the Higgs vacuum expectation value (VEV) which in turn induces shifts in the masses of fundamental SM particles.  For example, the shift in the electon mass is given by,
\begin{align}
m_e \approx m_e^\text{bare}\left(1 - \frac{\beta \rho_\phi^2}{2 m^2 m_H^2}\right),
\end{align}
where $m_e^\text{bare} \approx 0.5\,\text{MeV}$ is the unperturbed electron mass, $m_H$ is the Higgs mass, and $\rho_\phi$ is the expectation value of the boson star given by Eq. (\ref{eq:vphi}).  We note that this infers that the Higgs portal parameters must be fine-tuned \cite{Alonso-Alvarez:2019pfe,Kouvaris:2019nzd}.

The Higgs coupling constant is strenuously constrained from invsible Higgs decay, Big Bang Nucleosynthesis (BBN), fifth force experiments, and measurements of the electron and muon magnetic moment anomalies.  The rate for invsible Higgs decay is given by \cite{Kouvaris:2014uoa},
\begin{align}
\Gamma(h\rightarrow\phi\phi)\approx\frac{\beta^2 v_\text{EW}^2}{8 \pi m_H}\sqrt{1 - \left(\frac{2 m}{m_H}\right)^2},
\end{align}
where $v_\text{EW}$ is the electroweak VEV for $\beta = 0$.  Taking $m\ll m_H$, and using the CMS collaboration measurement of the branching fraction of invisible Higgs decays \cite{Sirunyan:2018owy} the upper constraint on $\beta$ is shown in Table \ref{tab:Higgs_consts}.

An upper constraint can also be placed on $\beta$ from constraints on the Fermi constant throughout the evolution of the universe.  Using the constraint on the change in the Fermi constant from the time of BBN to today \cite{Scherrer:1992na}, $\beta$ can be constrained from the ratio \cite{Alonso-Alvarez:2019pfe},
\begin{align}
\frac{G_{F}^\text{BBN}}{G_F^0} = \frac{1-\beta \rho_{\text{DM},0}^\text{avg}/(2 m^2 m_H^2)}{1-\beta \rho_{\text{DM,BBN}}^\text{avg}/(2 m^2 m_H^2)},
\end{align}
where $\rho_{\text{DM},0}^\text{avg} = 1.3 \, \text{keV}\,\text{cm}^{-3}$ is the average DM density of the universe today and $\rho_{\text{DM,BBN}}^\text{avg}$ as the time of BBN.  The resulting constraint on $\beta$ is shown in Table \ref{tab:Higgs_consts}.
\begin{table}[h!]
  \begin{center}
    \caption{Constraints on the Higgs coupling constant}
    \label{tab:Higgs_consts}
    \begin{tabular}{|c||c|c|}
	\hline
	& Invisible Higgs & BBN
	\\ \hline
	$\beta_\text{max}$ & $10^{-2}$ 
	& $2 \times 10^{-10} \left(\frac{m}{\mu \, \text{eV}}\right)^2 \left(\frac{1.3 \, \text{keV}\, \text{cm}^{-3}}{\rho_{\text{DM},0}^\text{avg}}\right)$ 
	\\\hline
    \end{tabular}
  \end{center}
\end{table}

\begin{figure}[t] 
\centering
\includegraphics[width=0.5\linewidth,height=250pt]{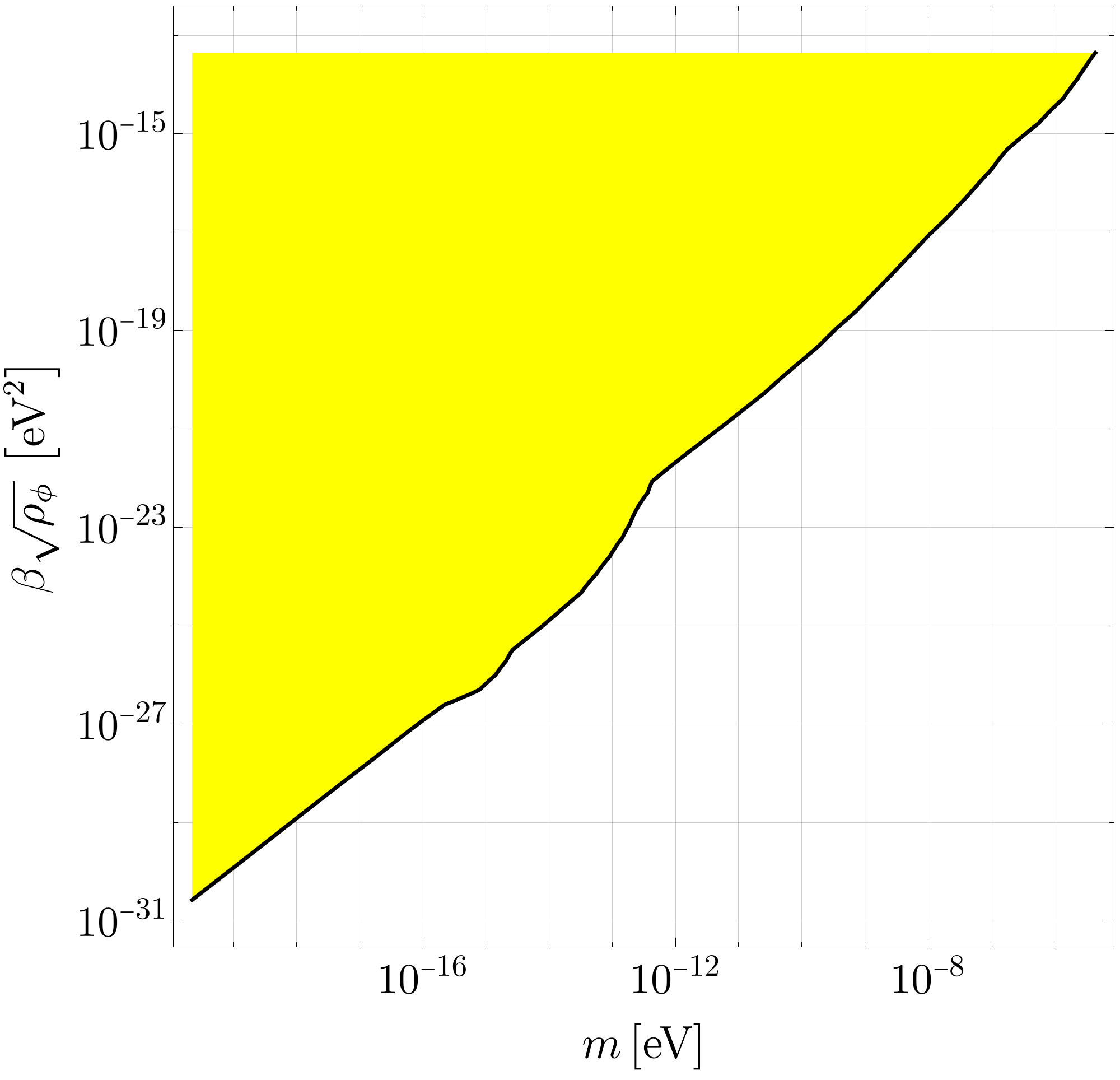} 
\caption{Constraints on $\beta \, \sqrt{\rho_\phi}$ from fifth force constraints \cite{Piazza:2010ye,Schlamminger:2007ht,Bertotti:2003rm,Kapner:2006si} given by Eq. (\ref{eq:ffconst}).  The plot has been reproduced from \cite{Derevianko:2013oaa}.} \label{fig:fifthforce}
\end{figure}

The Higgs coupling constant is constrained from fifth force experiments due to the constant presence of a nonzero density from the surrounding boson star.  The presence of $
\phi$ with mass $m$ induces a potential between two massive bodies given by \cite{Piazza:2010ye,Afach:2021pfd,PhysRevLett.123.091601},
\begin{align}
V(r) = -\frac{m_1 m_2}{r} \left(\frac{\alpha_{\phi N}}{M_P}\right)^2 e^{-mr},
\end{align}
where the coupling constant $\alpha_{\phi N}$ is given by,
\begin{align}
\alpha_{\phi N} = g_{h NN}\frac{\sqrt{2}M_P}{m_N} \epsilon.
\end{align}
Here, $m_N$ is the nucleon mass and $\epsilon$ the mixing parameter given by,
\begin{align}
\epsilon = \frac{\beta v_\text{EW}}{m_H^2}\sqrt{\frac{\rho_\phi}{2 m^2}}.
\end{align}
Using the value of $g_{hNN}$ derived in \cite{Afach:2021pfd,JUNGMAN1996195,HAIYANGCHENG1989347,GASSER1991252}, the parameter that is constrained from fifth force experiments \cite{Piazza:2010ye,Schlamminger:2007ht,Bertotti:2003rm,Kapner:2006si} is a function of the boson mass and is related to the Higgs portal coupling and boson star density as,
\begin{align}\label{eq:ffconst}
\alpha_{\phi N}^2(m) \sim 10^{29} \beta^2 \left(\frac{\rho_\phi/(2m^2)}{\text{GeV}^2}\right),
\end{align}
where $\rho_\phi$ is the density of the boson star given by Eq. (\ref{eq:vphi}).  Fig. \ref{fig:fifthforce} shows the constraints on $\beta \, \sqrt{\rho_\phi}$ for a given boson mass, where the relevant excluded parameter space was reproduced from \cite{Derevianko:2013oaa}.  We also ensure that the parameter space searched satisfies measured values of the muon and electron magnetic moments.  In this case, the Higgs portal will contribute to the muon and electron magnetic moments as \cite{Primack:1973iz,Bardeen:1972vi,Leveille:1977rc,Haber:1978jt,Carlson:1988dp,Zhou:2001ew},
\begin{align}\label{eq:mmconst}
\delta a_l = \frac{3}{32 \pi^2} Y_\text{eff}^2 = \frac{3}{32 \pi^2}\left(\frac{m_l \beta}{m_H^2}\right)^2 \frac{\rho_\phi}{2m^2},
\end{align}
where $m_l$ is the muon or electron mass and it is assumed that $m_l \gg m$.

\subsection{Photon portal}\label{sec:photon_portal}
We also consider the interaction between the DM and the photon through a photon portal \cite{Stadnik:2015kia},
\begin{align}
\mathcal{L} = ... + \frac{g}{4}|\phi|^2 F^2
\end{align}
where $g$ is a coupling constant that can be positive or negative.  This interaction induces a shift in the fine structure constant,
\begin{align}
\alpha \approx \alpha_0 \left(1 + g \frac{\rho_\phi}{2m^2}\right),
\end{align}
where $\alpha_0$ is the unperturbed fine structure constant and $\rho_\phi$ is given be Eq. (\ref{eq:vphi}).

The photon coupling constant is most strenuously constrained from supernova (SN) cooling and BBN \cite{Stadnik:2015kia,Stadnik:2015uka}.  These constraints are shown in Table \ref{tab:Photon_consts} where $g_\text{max}$ is the maximum photon coupling constant and $m$ is the DM mass.
\begin{table}[h!]
  \begin{center}
    \caption{Constraints on the Photon coupling constant}
    \label{tab:Photon_consts}
    \begin{tabular}{|c||c|c|}
	\hline
	& SN & BBN
	\\ \hline
	$g_\text{max} \, [\text{GeV}^{-2}]$ & $10^{-7}$ 
	& $8 \times 10^{-14} \left(\frac{m}{\mu\text{eV}}\right)^2$ 
	\\\hline
    \end{tabular}
  \end{center}
\end{table}
The BBN constraint can be derived from the constraint on the neutron-proton mass difference $Q_{np}$ where $Q_{np} \propto \alpha \Lambda_\text{QCD}$.  The shift in the neutron-proton mass difference due to the $\phi$ field is given by \cite{Stadnik:2015uka},
\begin{align}
\frac{\Delta (Q_{np}/T_F)}{Q_{np}/T_F} \approx 0.08 \frac{\Delta \alpha}{\alpha}\approx 0.0033
\end{align}
where $T_F \approx 0.8 \, \text{MeV}$ is the temperature at weak interaction freeze-out.  Using $\rho_\text{DM,BBN}^\text{avg} = \rho_\text{DM,0}^\text{avg}(1+z)^3$ at the time of BBN, the constraint on $g$ is found to be that in Table \ref{tab:Photon_consts}.

\section{Frequency shifts}\label{sec:freqshifts}
The shift in frequency of atomic clocks is given by \cite{Derevianko:2013oaa},
\begin{align}
\frac{\delta \omega}{\omega} = \frac{\delta V}{V},
\end{align}
where,
\begin{align}
V = \alpha^{K_\alpha}\left(\frac{m_q}{\Lambda_\text{QCD}}\right)^{K_q}\left(\frac{m_e}{m_p}\right)^{K_{e/p}},
\end{align}
and $\alpha$, $m_q$, $\Lambda_\text{QCD}$, $m_e$, and $m_p$ are the fine structure constant, quark mass, scale of QCD, electron mass, and proton mass, respectively.  The constants $K_\alpha$, $K_q$, and $K_{e/p}$ depend on the type of atomic clock in question. 

Taking a typical microwave atomic clock \cite{Derevianko:2013oaa}, $K_\alpha \simeq 2$, $K_q \sim -0.09$, and $K_{e/p} = 1$.  The shift in frequency due to the presence of a boson star that interacts with the SM through the Higgs portal is given by,
\begin{align}
\left(\frac{\delta \omega}{\omega}(r)\right)_\text{Higgs}\approx \frac{\delta m_e}{m_e} \approx \frac{\beta \rho_\phi(r)}{2 m^2 m_H^2},
\end{align}
where we show explicitly that the shift in frequency is a function of position within the boson star.  Here, we assume that the main contribution to the shift in frequency is due to the electron mass.  We neglect the change in mass of the quarks due to the fact that $|K_{e/p}| \gg |K_q|$ and we neglect the change in the proton mass due to the fact that most of the proton mass does not come from the quark masses.  

Taking a typical optical atomic clock, $K_{e/p}=K_q=0$ and $K_\alpha = 1$.  In this case, optical atomic clocks will not be sensitive to the DM interacting with the SM through the Higgs portal.  If the boson star is comprised of dark matter that interacts with the standard model through the photon portal, the shift in frequency is given by,
\begin{align}
\left(\frac{\delta \omega}{\omega}(r)\right)_\text{photon} \approx \frac{\delta \alpha}{\alpha} \approx g \frac{\rho_\phi(r)}{2m^2}.
\end{align}
Notice that both microwave and optical atomic clocks are sensitive to DM interacting with the SM through the photon portal.  However, optical atomic clocks will, in general, have better precision than microwave atomic clocks.

Consider two atomic clocks at positions $r_1$ and $r_2$ from the center of the boson star.  The shift in frequency between the two clocks is,
\begin{align}\label{eq:shift_diff}
\left(\frac{\delta \omega}{\omega}(r_1,r_2)\right)_\text{diff} = \frac{\delta \omega}{\omega}(r_1) - \frac{\delta \omega}{\omega}(r_2).
\end{align}
Scaling the positions as in Eq. (\ref{eq:scaling}), the frequency shifts at a scaled position $\tilde{r}_i$ for the Higgs and photon portals are given by,
\begin{align}
\left(\frac{\delta \omega}{\omega}(\tilde{r}_i)\right)_\text{Higgs} &\approx \frac{\beta}{m_H^2} \frac{\rho_\phi\left(\tilde{r}_i\right)}{2m^2}\label{eq:shift_Higgs_i};
\\
\left(\frac{\delta \omega}{\omega}(\tilde{r}_i)\right)_\text{Photon} &\approx g \frac{\rho_\phi\left(\tilde{r}_i\right)}{2m^2}\label{eq:shift_Photon_i}.
\end{align}

Microwave atomic clocks, which can probe both the Higgs and photon portals, can have precisions on the order of $10^{-16}$ \cite{Weyers_2018}, with a suggested improvement of $10^{-17}(T(K)/300)^2$ \cite{Han_2019}.  Optical atomic clocks, which can only probe the photon portal, can have precisions on the order of $10^{-18}$ \cite{PhysRevLett.104.070802,Bloom:2013uoa}, with a record precision of $10^{-19}$ \cite{PhysRevLett.120.103201}.  It has also been suggested that nuclear clocks, which can probe both the Higgs and photon portals, can reach precisions of $10^{-19}$ \cite{PhysRevLett.108.120802,Peik:2020cwm}.

\section{Results and conclusion}\label{sec:results}
We consider an atomic clock on the surface of the Earth and a second clock aboard the International Space Station (ISS).  We then transform the coordinates of the clocks in their respective lab frames to the heliocentric-ecliptic coordinates for the solar halo and to the geocentric-equatorial coordinates for the Earth halo \cite{McCabe:2013kea,Emken:2017qmp} (see Appendix \ref{sec:appA} for details).  The position of the ISS in the Earth's equatorial frame is given in \cite{Celestrak}.

\subsection{Solar halo}

We search the parameter space for frequency shifts of atomic clocks due to a boson star surrounding the Sun.  In particular, we search for parameters that give a difference in frequency shifts between the two clocks greater than some threshold frequency $\left(\frac{\delta \omega}{\omega}\right)_\text{diff, min}$ where the difference is given by Eq. (\ref{eq:shift_diff}).  We are interested in parameter space for which the solar centered boson star has a radius of at least $1 \text{AU}$.  For this case, we can safely take $\tilde{r}_d \gg 2 \tilde{r}_*$ where $\tilde{r}_d$ is given by Eq. (\ref{eq:rhosol}) and with $\tilde{r}_*$ and $n_*$ corresponding to the scaled radius and mass of the Sun given by Eq. (\ref{eq:scaling}).  For this regime (i.e. $\tilde{r}_d \gg 2 \tilde{r}_*$), we can safely use the external gravitational energy given by Eq. (\ref{eq:egapp1}).  We also impose the constraints on the boson star density at a given radius as shown in Table \ref{tab:bst_dens_consts}.  For the Higgs portal, we impose the constraints on the coupling constant $\beta$ from invisible Higgs decays and BBN as shown in Table \ref{tab:Higgs_consts}, from fifth force experiments (Eq.(\ref{eq:ffconst}) and Fig. \ref{fig:fifthforce}), and from the muon and electron anomalous magnetic moments (Eq. (\ref{eq:mmconst})).  For the photon portal, we impose constraints on the photon coupling constant $g$ from SN cooling and BBN as shown in Table \ref{tab:Photon_consts}.

We scan the available parameter space varying the self-coupling constant of the bosons within the range, $10^{-100} \leq \lambda \leq 1$, the mass of the boson within $10^{-31}\, \text{GeV} \leq m \leq 1 \, \text{GeV}$, and the scaled particle number of the boson star within $10^{-5}\, \left(n_c\right)_I \leq n \leq \left(n_c\right)_I$, where $\left(n_c\right)_I$ is the critical particle number given by Eq. (\ref{eq:ncrit}).  For the Higgs portal, we vary the Higgs coupling constant $\beta$ within $10^{-50} \leq \beta \leq 10^{-2}$, and for the photon portal, we vary the photon coupling constant $g$ within $10^{-50} \, \text{GeV}^{-2} \leq g \leq 10^{-7} \, \text{GeV}^{-2}$.  As stated previously, we require that the boson star density, Higgs coupling constant, and photon coupling constant satisfy all the necessary constraints.

The difference in frquency shifts between the two clocks is given by Eq. (\ref{eq:shift_diff}) with the frequency shift at each clock for the Higgs portal given by Eq. (\ref{eq:shift_Higgs_i}) and for the photon portal given by Eq. (\ref{eq:shift_Photon_i}) assuming for both $\left(\tilde{r}_d\right)_I$ as the minimum energy solution.  As an example, we take clock 1 to be on the surface of the Earth at a latitude $\Phi_1=0$ and longitude $\lambda_1=0$ and clock 2 aboard the ISS.  We take parameters for clock 2 corresponding to orbits 3531-3603 found in \cite{Celestrak} which were all measured in the month of July 2020.  For this analysis, we find no possible parameter space that can be measured with atomic clocks within the near future, as for both the Higgs and photon portals the maximum frequency shift possible is $\mathcal{O}(10^{-27})$.  This is mainly due the density constraints given in Table \ref{tab:bst_dens_consts}, resulting in boson star densities that can only induce frequency shifts several orders of magnitude smaller than the best precision of current optical and microwave atomic clocks.

\subsection{Earth halo}
For the Earth bound boson star, we find much more promising results.  We take the same example as discussed above, but for a boson star surrounding the Earth, which has a mass subject to the constraint given by Eq. (\ref{eq:massconst_earth}).  For this case, we assume the scaled radius of the boson star can be anywhere between $\tilde{r}_d \gg \tilde{r}_*$ and $\tilde{r}_d = \tilde{r}_*$ where $\tilde{r}_*$ is the scaled radius of the Earth.  In this case, we take two separate scans: one corresponding to $\tilde{r}_d \geq 2 \tilde{r}_*$ with the external gravitational energy given by Eq. (\ref{eq:egapp1}) and the minimum energy solution given by Eq. (\ref{eq:rhosol}); the other corresponding to $\tilde{r}_* \leq \tilde{r}_d < 2 \tilde{r}_*$ with the external gravitational energy given by Eq. (\ref{eq:egapp2}) and the minimum energy solution given by $\left(\tilde{r}_d\right)_{II}$.

\begin{figure}[t] 
\centering
\includegraphics[width=0.45\linewidth,height=500pt]{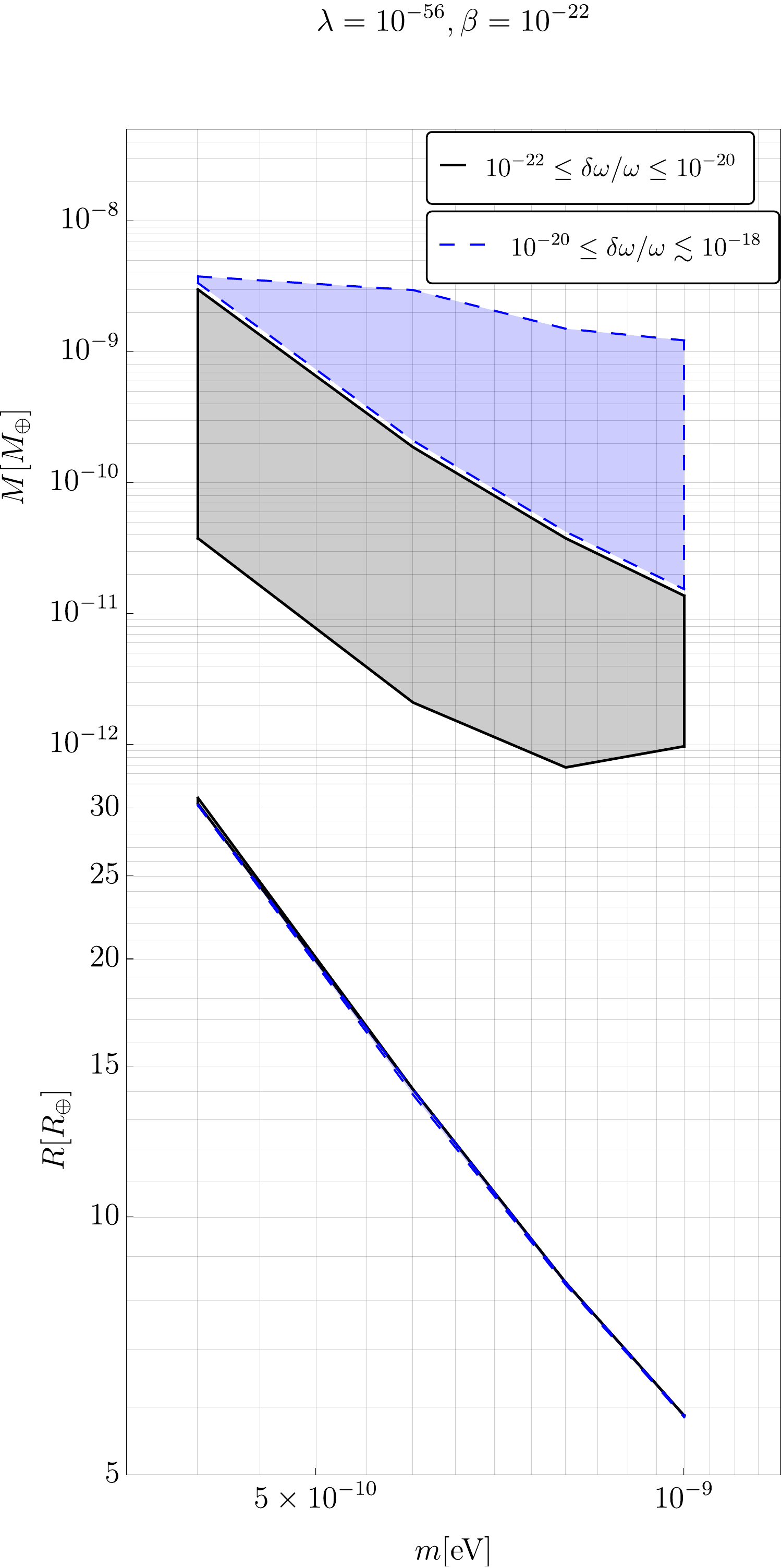} 
\,
\includegraphics[width=0.45\linewidth,height=500pt]{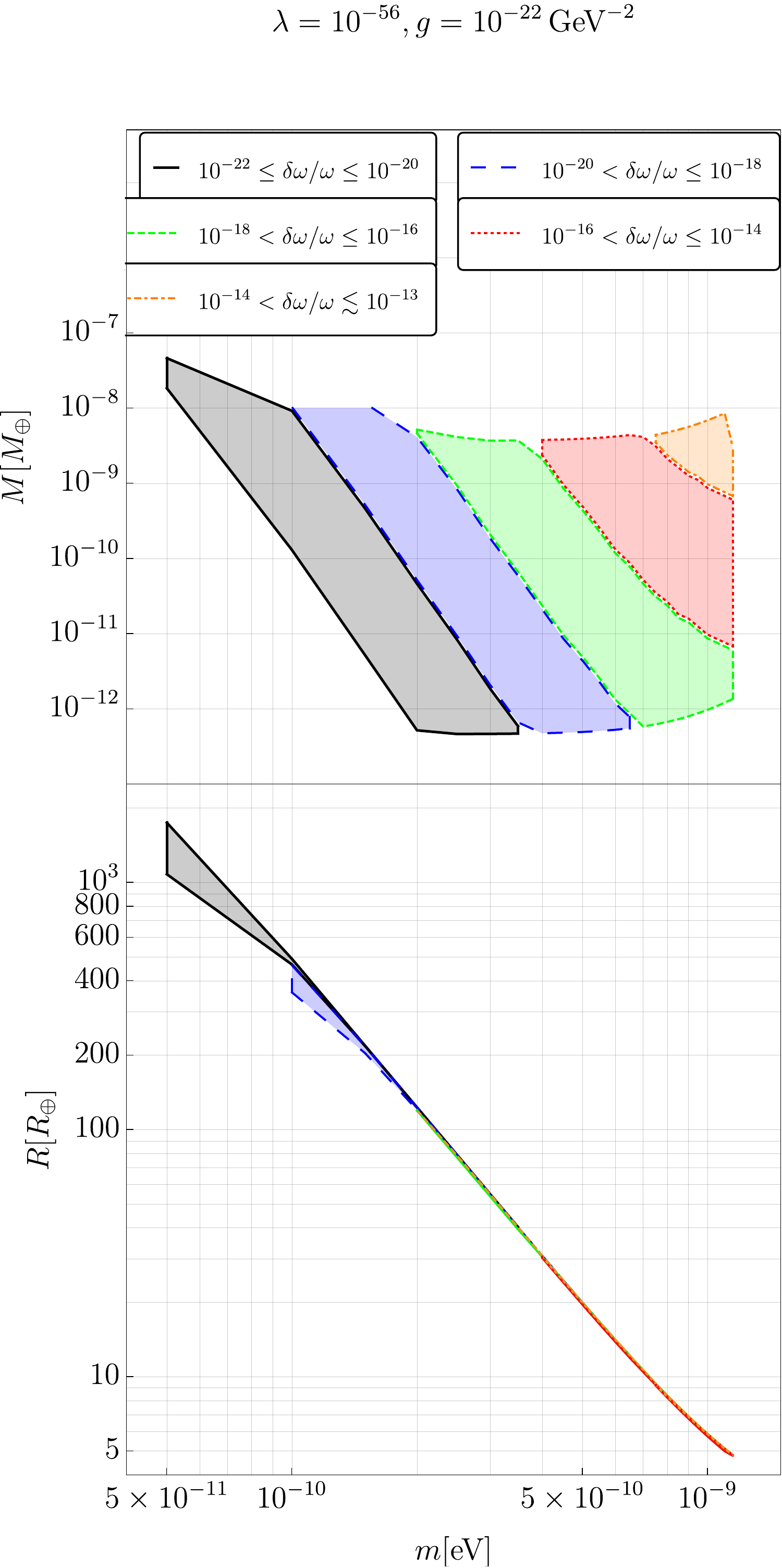} 
\caption{The total mass (top panel) and radius (bottom panel) of the boson star in units of the Earth mass and the Earth radius, respectively.  Clock 1 is assumed to be on the surface of the Earth at a latitude $\Phi_1=0$ and longitude $\lambda_1=0$ and clock 2 aboard the ISS.  The parameters for clock 2 are found in \cite{Celestrak} corresponding to orbit 3535 which were measured on July, 8, 2020 at 18:23:30 GMT.  \textbf{Left: Higgs portal}  The available parameter space taking $\lambda = 10^{-56}$ and $\beta = 10^{-22}$.  The black (thick) shaded region in each panel corresponds to a difference in frequency shift between the two clocks of $10^{-22} \leq \left(\delta \omega/\omega\right)_\text{diff} \leq 10^{-20}$ and the blue (large dashed) shaded region to $10^{-20} \leq \left(\delta \omega/\omega\right)_\text{diff} \lesssim 10^{-18}$.  \textbf{Right:  Photon portal}  The available parameter space taking $\lambda = 10^{-56}$ and $g = 10^{-22} \, \text{GeV}^{-2}$.  The black (thick) shaded region in each panel corresponds to a difference in frequency shift between the two clocks of $10^{-22} \leq \left(\delta \omega/\omega\right)_\text{diff} \leq 10^{-20}$, the blue (large dashed) shaded region to $10^{-20} \leq \left(\delta \omega/\omega\right)_\text{diff} < 10^{-18}$, the green (small dashed) shaded region to $10^{-18} \leq \left(\delta \omega/\omega\right)_\text{diff} < 10^{-16}$,  the red (dotted) shaded region to $10^{-16} \leq \left(\delta \omega/\omega\right)_\text{diff} < 10^{-14}$, and  the orange (dot-dashed) shaded region to $10^{-14} \leq \left(\delta \omega/\omega\right)_\text{diff} \lesssim 10^{-13}$.} \label{fig:paramspace}
\end{figure}

Fig. \ref{fig:paramspace} shows the available parameter space for the Higgs (left panel) and photon (right panel) portal for chosen values of $\lambda = 10^{-56}$, $\beta = 10^{-22}$ (left panel), and $g = 10^{-22} \, \text{GeV}^{-2}$ (right panel).  One can see that there is some small parameter space which future microwave atomic clocks can potentially probe for the Higgs portal, as the largest possible frequency shift that can be induced is $\sim 10^{-18}$.  However, the photon portal gives a much larger available parameter space and can potentially be probed with current optical atomic clocks as the largest possible frequency shift that can be induced is $\sim 10^{-13}$.

The various constraints we have aforementioned affect the parameter
space that can be probed in our scenario by use of atomic clocks. For
example, the parameter space in the upper panel is limited in small
boson masses (this is the left vertical side of the plot) due to the BBN
constraints  from Tables \ref{tab:Higgs_consts} and \ref{tab:Photon_consts}. On the right side (i.e., large boson masses), the parameter space is sharply limited due to our demand
that the radius of the boson star exceeds that of the Earth. The parameter
space is limited from below due to the atomic clock sensitivity, since
there is an experimental threshold on the fractional frequency
difference that can be detected.  Although atomic clocks do not exlude the
parameter space from above, it is nevertheless
restricted (i.e., there is an upper halo mass) for two
reasons: a) the constraint, Eq. (\ref{eq:massconst_earth}), that sets a limit on the maximum
allowed mass included in the vicinity of the Earth affecting the upper
right corner of the plot, and b) the maximum halo allowed mass provided
hydrodynamic stability. This constraint determines the boundary of the
constrained parameter space at the top of the upper panels.  Also note that while the total boson star masses are significantly sensitive to change in particle number of the boson star, the total boson star radii are not.  This can be seen by Eq. (\ref{eq:rhosol}) where $n \ll n_*$, and hence, $\tilde{r}_d$ does not depend significantly on $n$.

A comment is in order here. The uncertainty principle $\delta\omega \delta t >1$ sets a lower bound in $\delta t$. For part of our probed space (where $\delta \omega$ can be very small), $\delta t$ can become significantly large (e.g. of the order of a month) which practically means that the difference between two clocks cannot show up in general in a time interval smaller than what the uncertainty principle dictates. Therefore to probe extreme cases like these, stable atomic clocks which do not de-synchronize by random noise in intervals shorter than $\delta t$ are preferable. Atomic clocks with undisrupted synchronisation for periods of several months do exist making it possible to detect the hereby studied effect, since the latter is going to cause changes between the clocks within time intervals at least a few times shorter than what random noise can cause. One could imagine that our proposed effect can possibly be detected even with clocks that require more frequent synchronisation than $\delta t$. In that case, although the change in clocks will be hidden within random changes due to noise, given enough time and thus statistics, it could in principle be possible to identify the effect since it will always cause the same change between the clocks as opposed to the  random one due to noise.

In summary, we have entertained the idea that bosonic dark matter halos surround either the Sun or the Earth.  We have discussed the constraints that exclude parts of the potential parameter space, and have showed the remaining space that can potentially be probed by atomic clocks.  We have assumed either a Higgs portal or a photon portal and that the method of detection is a frequency comparison between a clock on Earth and a clock in space.  We have found that bosonic dark matter halos surrounding the Sun cannot be probed with atomic clocks in the near future.  However, those surrounding the Earth can potentially be probed with current and future atomic clocks.

\clearpage

\begin{acknowledgments}
L. S. and L. C. R. W. thank Brian Hemingway for useful discussions pertaining to atomic clocks and Joshua Eby for discussions pertaining to halo constraints.  L. S. and L. C. R. W. thank the University of Cincinnati Office of Research Faculty Bridge Program for funding through the Faculty Bridge Grant.  L. S. also thanks the Department of Physics at the University of Cincinnati for financial support in the form of the Violet M. Diller Fellowship.
\end{acknowledgments}

\clearpage

\appendix

\section{Astrophysical coordinates}\label{sec:appA}

Here, we discuss the calculations needed to convert all lab frames to the heliocentric-ecliptic frame or to the geocentric-equatorial frame \cite{McCabe:2013kea,Emken:2017qmp}.  All times considered will be taken relative to J2000.0 (01.01.2000 12:00 GMT).  The number of fractional days for a given date $D.M.Y$ and time $h:m:s$ (UT) relative to J2000.0 is,
\begin{widetext}
\begin{align}\label{eq:nJ2000}
n_{J2000.0} = \left[365.25 \bar{Y}\right] + \left[30.61\left(\bar{M} + 1 \right)\right] + D + \frac{h}{24} + \frac{m}{24 \times 60} + \frac{s}{24 \times 60^2} - 730563.5,
\end{align}
\end{widetext}
where
\begin{align}
\bar{Y} &= 
\begin{cases}
Y - 1 & \text{if} \quad M=1,2
\\
Y & \text{if} \quad M >2
\end{cases}
\nonumber\\
\bar{M} &= 
\begin{cases}
M + 12 & \text{if} \quad M=1,2
\\
M & \text{if} \quad M >2
\end{cases},
\end{align}
and $\left[...\right]$ is the floor function.  The epoch is then defined as,
\begin{align}\label{eq:TJ2000}
T_{J2000.0} \equiv \frac{n_{J2000.0}}{36525}.
\end{align}
The position of an atomic clock will depend on the Local Apparent Sidereal Time (LAST) which is the time since the local meridian passed the vernal equinox (\Aries) given by,
\begin{align}
\text{LAST}(\lambda) = \text{GAST} + \frac{\lambda}{360^o} \, 86400 \, \text{s},
\end{align}
where $\lambda$ is the longitude of the clock and the Greenwich Apparent Sidereal Time (GAST) is,
\begin{align}
\text{GAST} = \text{GMST} + E_e\left(T_{J2000.0}\right).
\end{align}
The equation of equinoxes is,
\begin{widetext}
\begin{align}
E_e\left(T_{J2000.0}\right) \approx \Delta \psi \cos \epsilon_A + 1.76 \times 10^{-4} \sin \Omega  \,\text{s} + 4 \times 10^{-6} \sin 2\Omega \, \text{s},
\end{align}
\end{widetext}
where
\begin{widetext}
\begin{align}
\Delta \psi &= - 1.1484 \sin \Omega \, \text{s} - 0.0864 \cos 2L \, \text{s}
\nonumber \\
\Omega &= 125.04455501^o - 0.05295376^o \, n_{J2000.0} + \mathcal{O}\left(T_{J2000.0}^2\right)
\nonumber \\
L &= 280.47^o - 0.98565^0 \,n_{J2000.0}+\mathcal{O}\left(T_{J2000.0}^2\right)
\nonumber \\
\epsilon_A &= 23.439279444^o - 0.01301021361^o T_{J2000.0} + \mathcal{O}\left(T_{J2000.0}^2\right).
\end{align}
\end{widetext}
Finally, the Greenwich Mean Sidereal Time (GMST) is,
\begin{widetext}
\begin{align}
\text{GMST} &= 86400\,\text{s} \left[0.7790572732640 + n_{J2000.0} \, \left(\text{mod} \, 1\right) + 0.00273781191135448\, n_{J2000.0} \right]
\nonumber \\
&+ 9.6707\times 10^{-4} \, \text{s} + 307.47710227\, T_{J2000.0}\,\text{s} + 0.092772113 \,T_{J2000.0}^2 \, \text{s} + \mathcal{O}\left(T_{J2000.0}^3\right).
\end{align}
\end{widetext}

The position of an atomic clock in the lab frame at a latitude $\Phi$ and longitude $\lambda$ can be described in the heliocentric-ecliptic coordinate system through a rotation.  A summary of the coordinate systems used is as follows:
\begin{enumerate}
\item lab frame (lab):  atomic clock at the origin, $x$-axis towards east; $y$-axis towards north; $z$-axis towards sky
\item geocentric-equatorial frame (g-eq):  center of Earth at the origin; $x$-axis towards \Aries; $z$-axis towards Earth north pole; $x$- and $y$-axes span the equatorial plane
\item geocentric-eclipctic (g-ecl):  center of Earth at the origin; $x$-axis towards \Aries; $z$-axis towards ecliptic north pole; $x$- and $y$-axes span the ecliptic plane
\item heliocentric-ecliptic (h-ecl):  center of Sun at the origin; $x$-axis towards \Aries; $z$-axis towards ecliptic north pole; $x$- and $y$- axies span the eclipctic plane
\end{enumerate}

\begin{figure}[t] 
\centering
\includegraphics[width=0.4\linewidth,height=150pt]{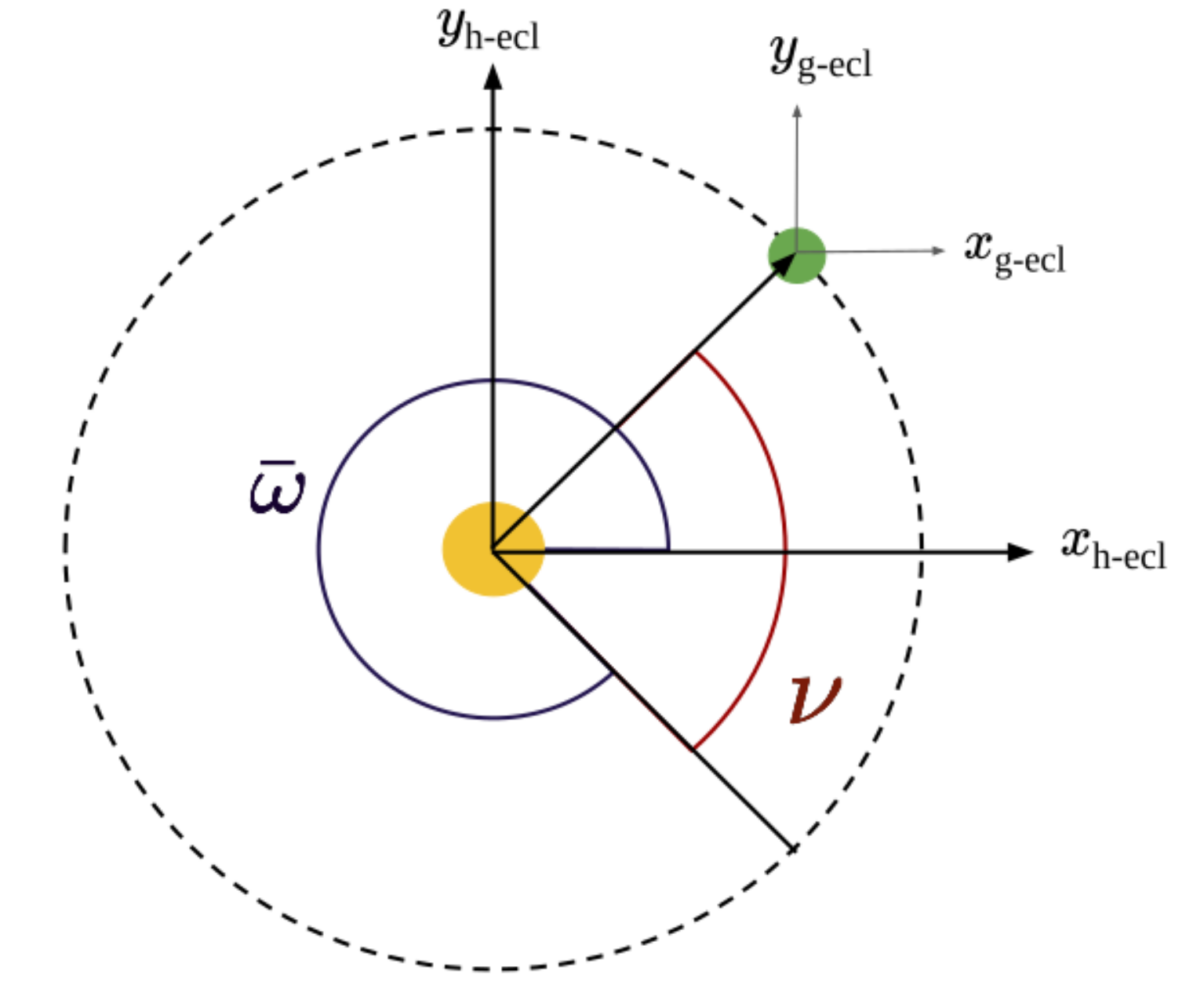} 
\\[\abovecaptionskip]
\caption{The heliocentric-ecliptic (h-ecl) and geocentric-ecliptic (g-ecl) coordinate systems.  Figure adapted from \cite{Emken:2017qmp}.} \label{fig:coord_sys}
\end{figure}
The rotation needed to go from the lab to g-ecl is then,
\begin{align}\label{eq:xgecl}
\bold{x}^\text{g-ecl} = \left(\mathcal{N}_\text{g-ecl}^\text{g-eq}\right)^{-1} \mathcal{N}_\text{lab}^\text{g-eq} \bold{x}^\text{lab}.
\end{align}
Here, $\mathcal{N}_\text{lab}^\text{g-eq} $ is the rotation matrix that takes coordinates in lab to coordinates in g-eq,
\begin{align}
\mathcal{N}_\text{lab}^\text{g-eq}  = 
\begin{pmatrix}
-\sin\phi & -\cos\theta\cos\phi & \sin\theta\cos\phi
\\
\cos\phi & -\cos\theta\sin\phi & \sin\theta\sin\phi
\\
0 & \sin\theta & \cos\theta
\end{pmatrix},
\end{align}
where $\theta = 90^o - \Phi$, $\phi(t) = \omega\,\text{LAST}(\lambda)$, and $\omega$ is the rotational velocity of the clock.  For the case of a clock 1 on the surface of the Earth, this rotational velocity is $\omega_1 = 2\pi/(86400\,\text{s})$ and the radius vector in the equatorial frame is,
\begin{align}
\bold{x}^\text{lab}(t) = 1 \, \text{AU}\begin{pmatrix}
\sin \theta \cos \phi(t)
\\
\sin \theta \sin \phi(t)
\\
\cos \theta
\end{pmatrix}.
\end{align}
The rotation matrix $\mathcal{N}_\text{g-ecl}^\text{g-eq}$ that takes coordinates in g-ecl to coordinates in g-eq is,
\begin{align}\label{eq:gecl}
\mathcal{N}_\text{g-ecl}^\text{g-eq} =
\begin{pmatrix}
1 & 0 & 0
\\
0 & \cos\epsilon & -\sin\epsilon
\\
0 &\sin\epsilon &\cos\epsilon
\end{pmatrix},
\end{align}
where $\epsilon = 23.4393^o - 0.0130^o \, T_{\text{J}2000.0}$, where $T_{\text{J}2000.0}$, and $n_{\text{J}2000.0}$ are given by Eqs. (\ref{eq:TJ2000}) and (\ref{eq:nJ2000}), respectively.  Once the position of the atomic clock in g-ecl coordinates is found from Eq. (\ref{eq:gecl}), the rotation to go from coordinates in g-ecl to coordinates in h-ecl is,
\begin{align}
\bold{x}^\text{h-ecl}(t) =
\begin{pmatrix}
r\cos l +x^\text{g-ecl}
\\
r \sin l +y^\text{g-ecl}
\\
z^\text{g-ecl}
\end{pmatrix},
\end{align}
where the Earth-Sun radius is,
\begin{align}
r = \frac{a(1 - e^2)}{1+e\cos \nu},
\end{align}
where $a = 1\,\text{AU}$, $e = 0.01671$, and $\nu$ is the true anomaly given by,
\begin{align}\label{eq:true_anomaly}
\nu = g+2 e \sin g + \frac{5}{4}e^2 \sin 2g + \mathcal{O}(e^3),
\end{align}
where $g = 357.528^o + 0.9856003^0 \,n_{J2000}$ is the mean anomaly, and the ecliptic longitude is,
\begin{align}\label{eq:ecl_long}
l = \bar{\omega} + \nu,
\end{align}
where
\begin{align}
\bar{\omega} = 282.932^o + 0.0000471^o \,n_{J2000}.
\end{align}

The orbital position of the International Space Station in the g-eq frame can be found from \cite{Celestrak},
\begin{align}
\bold{x}^\text{ISS,g-eq} = \begin{pmatrix}
r_\text{ISS} \cos l_\text{ISS} \cos i_\text{ISS}
\\
r_\text{ISS} \sin l_\text{ISS} \cos i_\text{ISS}
\\
r_\text{ISS} \sin i_\text{ISS}
\end{pmatrix},
\end{align}
where $r_\text{ISS}$ is the Earth-ISS radius given by,
\begin{align}
r^\text{ISS} = \frac{a_\text{ISS}(1 - e_\text{ISS}^2)}{1 + e_\text{ISS}\cos \nu_\text{ISS}}.
\end{align}
The true anomaly of the ISS is given by Eq. (\ref{eq:true_anomaly}) with a mean anomaly of $g_\text{ISS}$ and an orbital eccentricity $e_\text{ISS}$, while the mean orbital radius is $a_\text{ISS}$.  The equatorial longitude is given by Eq. (\ref{eq:ecl_long}) with the argument of periapsis $\bar{\omega}_\text{ISS}$, and the equatorial latitude  is $i_\text{ISS}$.  All measurements can be found from \cite{Celestrak}.

\bibliography{atom_clock}

\providecommand{\href}[2]{#2}\begingroup\raggedright\begin{thebibliography}{100}

\bibitem{Aghanim:2018eyx}
{\scshape Planck} collaboration, \emph{{Planck 2018 results. VI. Cosmological
  parameters}},
  \href{https://doi.org/10.1051/0004-6361/201833910}{\emph{Astron. Astrophys.}
  {\bfseries 641} (2020) A6}
  [\href{https://arxiv.org/abs/1807.06209}{{\ttfamily 1807.06209}}].

\bibitem{Frampton_2009}
P.~H. Frampton, \emph{Identification of all dark matter as black holes},
  \href{https://doi.org/10.1088/1475-7516/2009/10/016}{\emph{Journal of
  Cosmology and Astroparticle Physics} {\bfseries 2009} (2009) 016}.

\bibitem{Frampton_2010}
P.~H. Frampton, M.~Kawasaki, F.~Takahashi and T.~T. Yanagida, \emph{Primordial
  black holes as all dark matter},
  \href{https://doi.org/10.1088/1475-7516/2010/04/023}{\emph{Journal of
  Cosmology and Astroparticle Physics} {\bfseries 2010} (2010) 023}.

\bibitem{Garc_a_Bellido_2017}
J.~Garc{\'{\i}}a-Bellido, \emph{Massive primordial black holes as dark matter
  and their detection with gravitational waves},
  \href{https://doi.org/10.1088/1742-6596/840/1/012032}{\emph{Journal of
  Physics: Conference Series} {\bfseries 840} (2017) 012032}.

\bibitem{Peccei:1977hh}
R.~D. Peccei and H.~R. Quinn, \emph{{CP Conservation in the Presence of
  Instantons}}, \href{https://doi.org/10.1103/PhysRevLett.38.1440}{\emph{Phys.
  Rev. Lett.} {\bfseries 38} (1977) 1440}.

\bibitem{Peccei:1977ur}
R.~D. Peccei and H.~R. Quinn, \emph{{Constraints Imposed by CP Conservation in
  the Presence of Instantons}},
  \href{https://doi.org/10.1103/PhysRevD.16.1791}{\emph{Phys. Rev. D}
  {\bfseries 16} (1977) 1791}.

\bibitem{Weinberg:1977ma}
S.~Weinberg, \emph{{A New Light Boson?}},
  \href{https://doi.org/10.1103/PhysRevLett.40.223}{\emph{Phys. Rev. Lett.}
  {\bfseries 40} (1978) 223}.

\bibitem{Wilczek:1977pj}
F.~Wilczek, \emph{{Problem of Strong $P$ and $T$ Invariance in the Presence of
  Instantons}}, \href{https://doi.org/10.1103/PhysRevLett.40.279}{\emph{Phys.
  Rev. Lett.} {\bfseries 40} (1978) 279}.

\bibitem{DINE1981199}
M.~Dine, W.~Fischler and M.~Srednicki, \emph{A simple solution to the strong cp
  problem with a harmless axion},
  \href{https://doi.org/https://doi.org/10.1016/0370-2693(81)90590-6}{\emph{Physics
  Letters B} {\bfseries 104} (1981) 199}.

\bibitem{Kim:1979if}
J.~E. Kim, \emph{{Weak Interaction Singlet and Strong CP Invariance}},
  \href{https://doi.org/10.1103/PhysRevLett.43.103}{\emph{Phys. Rev. Lett.}
  {\bfseries 43} (1979) 103}.

\bibitem{SHIFMAN1980493}
M.~Shifman, A.~Vainshtein and V.~Zakharov, \emph{Can confinement ensure natural
  cp invariance of strong interactions?},
  \href{https://doi.org/https://doi.org/10.1016/0550-3213(80)90209-6}{\emph{Nuclear
  Physics B} {\bfseries 166} (1980) 493}.

\bibitem{Turner:1983he}
M.~S. Turner, \emph{{Coherent Scalar Field Oscillations in an Expanding
  Universe}}, \href{https://doi.org/10.1103/PhysRevD.28.1243}{\emph{Phys. Rev.
  D} {\bfseries 28} (1983) 1243}.

\bibitem{Press:1989id}
W.~H. Press, B.~S. Ryden and D.~N. Spergel, \emph{{Single Mechanism for
  Generating Large Scale Structure and Providing Dark Missing Matter}},
  \href{https://doi.org/10.1103/PhysRevLett.64.1084}{\emph{Phys. Rev. Lett.}
  {\bfseries 64} (1990) 1084}.

\bibitem{Sin:1992bg}
S.-J. Sin, \emph{{Late time cosmological phase transition and galactic halo as
  Bose liquid}}, \href{https://doi.org/10.1103/PhysRevD.50.3650}{\emph{Phys.
  Rev. D} {\bfseries 50} (1994) 3650}
  [\href{https://arxiv.org/abs/hep-ph/9205208}{{\ttfamily hep-ph/9205208}}].

\bibitem{Hu:2000ke}
W.~Hu, R.~Barkana and A.~Gruzinov, \emph{{Cold and fuzzy dark matter}},
  \href{https://doi.org/10.1103/PhysRevLett.85.1158}{\emph{Phys. Rev. Lett.}
  {\bfseries 85} (2000) 1158}
  [\href{https://arxiv.org/abs/astro-ph/0003365}{{\ttfamily
  astro-ph/0003365}}].

\bibitem{GOODMAN2000103}
J.~Goodman, \emph{Repulsive dark matter},
  \href{https://doi.org/https://doi.org/10.1016/S1384-1076(00)00015-4}{\emph{New
  Astronomy} {\bfseries 5} (2000) 103}.

\bibitem{Peebles_2000}
P.~J.~E. Peebles, \emph{Fluid dark matter},
  \href{https://doi.org/10.1086/312677}{\emph{The Astrophysical Journal}
  {\bfseries 534} (2000) L127}.

\bibitem{AMENDOLA2006192}
L.~Amendola and R.~Barbieri, \emph{Dark matter from an ultra-light
  pseudo-goldsone-boson},
  \href{https://doi.org/https://doi.org/10.1016/j.physletb.2006.08.069}{\emph{Physics
  Letters B} {\bfseries 642} (2006) 192}.

\bibitem{Arvanitaki:2009fg}
A.~Arvanitaki, S.~Dimopoulos, S.~Dubovsky, N.~Kaloper and J.~March-Russell,
  \emph{{String Axiverse}},
  \href{https://doi.org/10.1103/PhysRevD.81.123530}{\emph{Phys. Rev. D}
  {\bfseries 81} (2010) 123530}
  [\href{https://arxiv.org/abs/0905.4720}{{\ttfamily 0905.4720}}].

\bibitem{Li:2013nal}
B.~Li, T.~Rindler-Daller and P.~R. Shapiro, \emph{{Cosmological Constraints on
  Bose-Einstein-Condensed Scalar Field Dark Matter}},
  \href{https://doi.org/10.1103/PhysRevD.89.083536}{\emph{Phys. Rev. D}
  {\bfseries 89} (2014) 083536}
  [\href{https://arxiv.org/abs/1310.6061}{{\ttfamily 1310.6061}}].

\bibitem{MARSH20161}
D.~J. Marsh, \emph{Axion cosmology},
  \href{https://doi.org/https://doi.org/10.1016/j.physrep.2016.06.005}{\emph{Physics
  Reports} {\bfseries 643} (2016) 1}.

\bibitem{Hui:2016ltb}
L.~Hui, J.~P. Ostriker, S.~Tremaine and E.~Witten, \emph{{Ultralight scalars as
  cosmological dark matter}},
  \href{https://doi.org/10.1103/PhysRevD.95.043541}{\emph{Phys. Rev. D}
  {\bfseries 95} (2017) 043541}
  [\href{https://arxiv.org/abs/1610.08297}{{\ttfamily 1610.08297}}].

\bibitem{Lee:2017qve}
J.-W. Lee, \emph{{Brief History of Ultra-light Scalar Dark Matter Models}},
  \href{https://doi.org/10.1051/epjconf/201816806005}{\emph{EPJ Web Conf.}
  {\bfseries 168} (2018) 06005}
  [\href{https://arxiv.org/abs/1704.05057}{{\ttfamily 1704.05057}}].

\bibitem{Aprile:2018dbl}
{\scshape XENON} collaboration, \emph{{Dark Matter Search Results from a One
  Ton-Year Exposure of XENON1T}},
  \href{https://doi.org/10.1103/PhysRevLett.121.111302}{\emph{Phys. Rev. Lett.}
  {\bfseries 121} (2018) 111302}
  [\href{https://arxiv.org/abs/1805.12562}{{\ttfamily 1805.12562}}].

\bibitem{Slatyer:2017sev}
T.~R. Slatyer, \emph{{Indirect Detection of Dark Matter}},  in
  \emph{{Theoretical Advanced Study Institute in Elementary Particle Physics}:
  {Anticipating the Next Discoveries in Particle Physics}}, 10, 2017,
  \href{https://doi.org/10.1142/9789813233348_0005}{DOI}
  [\href{https://arxiv.org/abs/1710.05137}{{\ttfamily 1710.05137}}].

\bibitem{Petraki:2013wwa}
K.~Petraki and R.~R. Volkas, \emph{{Review of asymmetric dark matter}},
  \href{https://doi.org/10.1142/S0217751X13300287}{\emph{Int. J. Mod. Phys. A}
  {\bfseries 28} (2013) 1330028}
  [\href{https://arxiv.org/abs/1305.4939}{{\ttfamily 1305.4939}}].

\bibitem{Pollack_2015}
J.~Pollack, D.~N. Spergel and P.~J. Steinhardt, \emph{{SUPERMASSIVE} {BLACK}
  {HOLES} {FROM} {ULTRA}-{STRONGLY} {SELF}-{INTERACTING} {DARK} {MATTER}},
  \href{https://doi.org/10.1088/0004-637x/804/2/131}{\emph{The Astrophysical
  Journal} {\bfseries 804} (2015) 131}.

\bibitem{Chang_2019}
J.~H. Chang, D.~Egana-Ugrinovic, R.~Essig and C.~Kouvaris, \emph{Structure
  formation and exotic compact objects in a dissipative dark sector},
  \href{https://doi.org/10.1088/1475-7516/2019/03/036}{\emph{Journal of
  Cosmology and Astroparticle Physics} {\bfseries 2019} (2019) 036}.

\bibitem{Kouvaris:2015rea}
C.~Kouvaris and N.~G. Nielsen, \emph{{Asymmetric Dark Matter Stars}},
  \href{https://doi.org/10.1103/PhysRevD.92.063526}{\emph{Phys. Rev. D}
  {\bfseries 92} (2015) 063526}
  [\href{https://arxiv.org/abs/1507.00959}{{\ttfamily 1507.00959}}].

\bibitem{Alonso-Alvarez:2019pfe}
G.~Alonso-Álvarez, J.~Gehrlein, J.~Jaeckel and S.~Schenk, \emph{{Very Light
  Asymmetric Dark Matter}},
  \href{https://doi.org/10.1088/1475-7516/2019/09/003}{\emph{JCAP} {\bfseries
  09} (2019) 003} [\href{https://arxiv.org/abs/1906.00969}{{\ttfamily
  1906.00969}}].

\bibitem{Maselli:2019ubs}
A.~Maselli, C.~Kouvaris and K.~D. Kokkotas, \emph{{Photon spectrum of
  asymmetric dark stars}},
  \href{https://doi.org/10.1142/S0218271821500036}{\emph{Int. J. Mod. Phys. D}
  {\bfseries 30} (2021) 2150003}
  [\href{https://arxiv.org/abs/1905.05769}{{\ttfamily 1905.05769}}].

\bibitem{Derevianko:2013oaa}
A.~Derevianko and M.~Pospelov, \emph{{Hunting for topological dark matter with
  atomic clocks}}, \href{https://doi.org/10.1038/nphys3137}{\emph{Nature Phys.}
  {\bfseries 10} (2014) 933} [\href{https://arxiv.org/abs/1311.1244}{{\ttfamily
  1311.1244}}].

\bibitem{Roberts:2017hla}
B.~M. Roberts, G.~Blewitt, C.~Dailey, M.~Murphy, M.~Pospelov, A.~Rollings
  et~al., \emph{{Search for domain wall dark matter with atomic clocks on board
  global positioning system satellites}},
  \href{https://doi.org/10.1038/s41467-017-01440-4}{\emph{Nature Commun.}
  {\bfseries 8} (2017) 1195}
  [\href{https://arxiv.org/abs/1704.06844}{{\ttfamily 1704.06844}}].

\bibitem{Afach:2021pfd}
S.~Afach et~al., \emph{{Search for topological defect dark matter using the
  global network of optical magnetometers for exotic physics searches
  (GNOME)}},  \href{https://arxiv.org/abs/2102.13379}{{\ttfamily 2102.13379}}.

\bibitem{Arvanitaki:2014faa}
A.~Arvanitaki, J.~Huang and K.~Van~Tilburg, \emph{{Searching for dilaton dark
  matter with atomic clocks}},
  \href{https://doi.org/10.1103/PhysRevD.91.015015}{\emph{Phys. Rev. D}
  {\bfseries 91} (2015) 015015}
  [\href{https://arxiv.org/abs/1405.2925}{{\ttfamily 1405.2925}}].

\bibitem{Stadnik:2014tta}
Y.~V. Stadnik and V.~V. Flambaum, \emph{{Searching for dark matter and
  variation of fundamental constants with laser and maser interferometry}},
  \href{https://doi.org/10.1103/PhysRevLett.114.161301}{\emph{Phys. Rev. Lett.}
  {\bfseries 114} (2015) 161301}
  [\href{https://arxiv.org/abs/1412.7801}{{\ttfamily 1412.7801}}].

\bibitem{VanTilburg:2015oza}
K.~Van~Tilburg, N.~Leefer, L.~Bougas and D.~Budker, \emph{{Search for
  ultralight scalar dark matter with atomic spectroscopy}},
  \href{https://doi.org/10.1103/PhysRevLett.115.011802}{\emph{Phys. Rev. Lett.}
  {\bfseries 115} (2015) 011802}
  [\href{https://arxiv.org/abs/1503.06886}{{\ttfamily 1503.06886}}].

\bibitem{Hees:2016gop}
A.~Hees, J.~Gu\'ena, M.~Abgrall, S.~Bize and P.~Wolf, \emph{{Searching for an
  oscillating massive scalar field as a dark matter candidate using atomic
  hyperfine frequency comparisons}},
  \href{https://doi.org/10.1103/PhysRevLett.117.061301}{\emph{Phys. Rev. Lett.}
  {\bfseries 117} (2016) 061301}
  [\href{https://arxiv.org/abs/1604.08514}{{\ttfamily 1604.08514}}].

\bibitem{Hees:2018fpg}
A.~Hees, O.~Minazzoli, E.~Savalle, Y.~V. Stadnik and P.~Wolf, \emph{{Violation
  of the equivalence principle from light scalar dark matter}},
  \href{https://doi.org/10.1103/PhysRevD.98.064051}{\emph{Phys. Rev. D}
  {\bfseries 98} (2018) 064051}
  [\href{https://arxiv.org/abs/1807.04512}{{\ttfamily 1807.04512}}].

\bibitem{Roberts:2018agv}
B.~M. Roberts and A.~Derevianko, \emph{{Precision measurement noise asymmetry
  and its annual modulation as a dark matter signature}},
  \href{https://doi.org/10.3390/universe7030050}{\emph{Universe} {\bfseries 7}
  (2021) 50} [\href{https://arxiv.org/abs/1803.00617}{{\ttfamily 1803.00617}}].

\bibitem{Roberts:2018xqn}
B.~M. Roberts, G.~Blewitt, C.~Dailey and A.~Derevianko, \emph{{Search for
  transient ultralight dark matter signatures with networks of precision
  measurement devices using a Bayesian statistics method}},
  \href{https://doi.org/10.1103/PhysRevD.97.083009}{\emph{Phys. Rev. D}
  {\bfseries 97} (2018) 083009}
  [\href{https://arxiv.org/abs/1803.10264}{{\ttfamily 1803.10264}}].

\bibitem{Wolf:2018xlz}
P.~Wolf, R.~Alonso and D.~Blas, \emph{{Scattering of light dark matter in
  atomic clocks}},
  \href{https://doi.org/10.1103/PhysRevD.99.095019}{\emph{Phys. Rev. D}
  {\bfseries 99} (2019) 095019}
  [\href{https://arxiv.org/abs/1810.01632}{{\ttfamily 1810.01632}}].

\bibitem{Alonso:2018dxy}
R.~Alonso, D.~Blas and P.~Wolf, \emph{{Exploring the ultra-light to sub-MeV
  dark matter window with atomic clocks and co-magnetometers}},
  \href{https://doi.org/10.1007/JHEP07(2019)069}{\emph{JHEP} {\bfseries 07}
  (2019) 069} [\href{https://arxiv.org/abs/1810.00889}{{\ttfamily
  1810.00889}}].

\bibitem{Savalle:2019jsb}
E.~Savalle, B.~M. Roberts, F.~Frank, P.-E. Pottie, B.~T. McAllister, C.~Dailey
  et~al., \emph{{Novel approaches to dark-matter detection using space-time
  separated clocks}},  \href{https://arxiv.org/abs/1902.07192}{{\ttfamily
  1902.07192}}.

\bibitem{Kaup:1968zz}
D.~J. Kaup, \emph{{Klein-Gordon Geon}},
  \href{https://doi.org/10.1103/PhysRev.172.1331}{\emph{Phys. Rev.} {\bfseries
  172} (1968) 1331}.

\bibitem{Ruffini:1969qy}
R.~Ruffini and S.~Bonazzola, \emph{{Systems of selfgravitating particles in
  general relativity and the concept of an equation of state}},
  \href{https://doi.org/10.1103/PhysRev.187.1767}{\emph{Phys. Rev.} {\bfseries
  187} (1969) 1767}.

\bibitem{BREIT1984329}
J.~Breit, S.~Gupta and A.~Zaks, \emph{Cold bose stars},
  \href{https://doi.org/https://doi.org/10.1016/0370-2693(84)90764-0}{\emph{Physics
  Letters B} {\bfseries 140} (1984) 329}.

\bibitem{Colpi:1986ye}
M.~Colpi, S.~L. Shapiro and I.~Wasserman, \emph{{Boson Stars: Gravitational
  Equilibria of Selfinteracting Scalar Fields}},
  \href{https://doi.org/10.1103/PhysRevLett.57.2485}{\emph{Phys. Rev. Lett.}
  {\bfseries 57} (1986) 2485}.

\bibitem{Seidel:1990jh}
E.~Seidel and W.-M. Suen, \emph{{Dynamical Evolution of Boson Stars. 1.
  Perturbing the Ground State}},
  \href{https://doi.org/10.1103/PhysRevD.42.384}{\emph{Phys. Rev. D} {\bfseries
  42} (1990) 384}.

\bibitem{Friedberg:1986tq}
R.~Friedberg, T.~D. Lee and Y.~Pang, \emph{{Scalar Soliton Stars and Black
  Holes}}, \href{https://doi.org/10.1103/PhysRevD.35.3658}{\emph{Phys. Rev. D}
  {\bfseries 35} (1987) 3658}.

\bibitem{Seidel:1991zh}
E.~Seidel and W.~M. Suen, \emph{{Oscillating soliton stars}},
  \href{https://doi.org/10.1103/PhysRevLett.66.1659}{\emph{Phys. Rev. Lett.}
  {\bfseries 66} (1991) 1659}.

\bibitem{LEE1992251}
T.~Lee and Y.~Pang, \emph{Nontopological solitons},
  \href{https://doi.org/https://doi.org/10.1016/0370-1573(92)90064-7}{\emph{Physics
  Reports} {\bfseries 221} (1992) 251}.

\bibitem{Guzman:2006yc}
F.~S. Guzman and L.~A. Urena-Lopez, \emph{{Gravitational cooling of
  self-gravitating Bose-Condensates}},
  \href{https://doi.org/10.1086/504508}{\emph{Astrophys. J.} {\bfseries 645}
  (2006) 814} [\href{https://arxiv.org/abs/astro-ph/0603613}{{\ttfamily
  astro-ph/0603613}}].

\bibitem{PhysRevD.68.023511}
A.~Arbey, J.~Lesgourgues and P.~Salati, \emph{Galactic halos of fluid dark
  matter}, \href{https://doi.org/10.1103/PhysRevD.68.023511}{\emph{Phys. Rev.
  D} {\bfseries 68} (2003) 023511}.

\bibitem{PhysRevD.53.2236}
J.-w. Lee and I.-g. Koh, \emph{Galactic halos as boson stars},
  \href{https://doi.org/10.1103/PhysRevD.53.2236}{\emph{Phys. Rev. D}
  {\bfseries 53} (1996) 2236}.

\bibitem{Matos:2007zza}
T.~Matos and L.~A. Urena-Lopez, \emph{{Flat rotation curves in scalar field
  galaxy halos}}, \href{https://doi.org/10.1007/s10714-007-0470-y}{\emph{Gen.
  Rel. Grav.} {\bfseries 39} (2007) 1279}.

\bibitem{Bernal:2009zy}
A.~Bernal, J.~Barranco, D.~Alic and C.~Palenzuela, \emph{{Multi-state Boson
  Stars}}, \href{https://doi.org/10.1103/PhysRevD.81.044031}{\emph{Phys. Rev.
  D} {\bfseries 81} (2010) 044031}
  [\href{https://arxiv.org/abs/0908.2435}{{\ttfamily 0908.2435}}].

\bibitem{UrenaLopez:2010ur}
L.~A. Urena-Lopez and A.~Bernal, \emph{{Bosonic gas as a Galactic Dark Matter
  Halo}}, \href{https://doi.org/10.1103/PhysRevD.82.123535}{\emph{Phys. Rev. D}
  {\bfseries 82} (2010) 123535}
  [\href{https://arxiv.org/abs/1008.1231}{{\ttfamily 1008.1231}}].

\bibitem{Chavanis:2011zi}
P.-H. Chavanis, \emph{{Mass-radius relation of Newtonian self-gravitating
  Bose-Einstein condensates with short-range interactions: I. Analytical
  results}}, \href{https://doi.org/10.1103/PhysRevD.84.043531}{\emph{Phys. Rev.
  D} {\bfseries 84} (2011) 043531}
  [\href{https://arxiv.org/abs/1103.2050}{{\ttfamily 1103.2050}}].

\bibitem{Chavanis:2011zm}
P.~H. Chavanis and L.~Delfini, \emph{{Mass-radius relation of Newtonian
  self-gravitating Bose-Einstein condensates with short-range interactions: II.
  Numerical results}},
  \href{https://doi.org/10.1103/PhysRevD.84.043532}{\emph{Phys. Rev. D}
  {\bfseries 84} (2011) 043532}
  [\href{https://arxiv.org/abs/1103.2054}{{\ttfamily 1103.2054}}].

\bibitem{doi:10.1142/S0217732316500905}
J.~Eby, P.~Suranyi and L.~C.~R. Wijewardhana, \emph{The lifetime of axion
  stars}, \href{https://doi.org/10.1142/S0217732316500905}{\emph{Modern Physics
  Letters A} {\bfseries 31} (2016) 1650090}
  [\href{https://arxiv.org/abs/https://doi.org/10.1142/S0217732316500905}{{\ttfamily
  https://doi.org/10.1142/S0217732316500905}}].

\bibitem{Eby:2016cnq}
J.~Eby, M.~Leembruggen, P.~Suranyi and L.~C.~R. Wijewardhana, \emph{{Collapse
  of Axion Stars}}, \href{https://doi.org/10.1007/JHEP12(2016)066}{\emph{JHEP}
  {\bfseries 12} (2016) 066}
  [\href{https://arxiv.org/abs/1608.06911}{{\ttfamily 1608.06911}}].

\bibitem{Eby:2017azn}
J.~Eby, M.~Ma, P.~Suranyi and L.~C.~R. Wijewardhana, \emph{{Decay of Ultralight
  Axion Condensates}},
  \href{https://doi.org/10.1007/JHEP01(2018)066}{\emph{JHEP} {\bfseries 01}
  (2018) 066} [\href{https://arxiv.org/abs/1705.05385}{{\ttfamily
  1705.05385}}].

\bibitem{VISINELLI201864}
L.~Visinelli, S.~Baum, J.~Redondo, K.~Freese and F.~Wilczek, \emph{Dilute and
  dense axion stars},
  \href{https://doi.org/https://doi.org/10.1016/j.physletb.2017.12.010}{\emph{Physics
  Letters B} {\bfseries 777} (2018) 64}.

\bibitem{Levkov:2018kau}
D.~G. Levkov, A.~G. Panin and I.~I. Tkachev, \emph{{Gravitational Bose-Einstein
  condensation in the kinetic regime}},
  \href{https://doi.org/10.1103/PhysRevLett.121.151301}{\emph{Phys. Rev. Lett.}
  {\bfseries 121} (2018) 151301}
  [\href{https://arxiv.org/abs/1804.05857}{{\ttfamily 1804.05857}}].

\bibitem{Lin:2018whl}
S.-C. Lin, H.-Y. Schive, S.-K. Wong and T.~Chiueh, \emph{{Self-consistent
  construction of virialized wave dark matter halos}},
  \href{https://doi.org/10.1103/PhysRevD.97.103523}{\emph{Phys. Rev. D}
  {\bfseries 97} (2018) 103523}
  [\href{https://arxiv.org/abs/1801.02320}{{\ttfamily 1801.02320}}].

\bibitem{Guzman:2019gqc}
F.~S. Guzm\'an and L.~A. Ure\~na L\'opez, \emph{{Gravitational atoms: General
  framework for the construction of multistate axially symmetric solutions of
  the Schr\"odinger-Poisson system}},
  \href{https://doi.org/10.1103/PhysRevD.101.081302}{\emph{Phys. Rev. D}
  {\bfseries 101} (2020) 081302}
  [\href{https://arxiv.org/abs/1912.10585}{{\ttfamily 1912.10585}}].

\bibitem{Braaten:2019knj}
E.~Braaten and H.~Zhang, \emph{{Colloquium : The physics of axion stars}},
  \href{https://doi.org/10.1103/RevModPhys.91.041002}{\emph{Rev. Mod. Phys.}
  {\bfseries 91} (2019) 041002}.

\bibitem{sym12010025}
H.~Zhang, \emph{Axion stars},
  \href{https://doi.org/10.3390/sym12010025}{\emph{Symmetry} {\bfseries 12}
  (2020) }.

\bibitem{Eby:2019ntd}
J.~Eby, M.~Leembruggen, L.~Street, P.~Suranyi and L.~C.~R. Wijewardhana,
  \emph{{Global view of QCD axion stars}},
  \href{https://doi.org/10.1103/PhysRevD.100.063002}{\emph{Phys. Rev. D}
  {\bfseries 100} (2019) 063002}
  [\href{https://arxiv.org/abs/1905.00981}{{\ttfamily 1905.00981}}].

\bibitem{Eggemeier:2019jsu}
B.~Eggemeier and J.~C. Niemeyer, \emph{{Formation and mass growth of axion
  stars in axion miniclusters}},
  \href{https://doi.org/10.1103/PhysRevD.100.063528}{\emph{Phys. Rev. D}
  {\bfseries 100} (2019) 063528}
  [\href{https://arxiv.org/abs/1906.01348}{{\ttfamily 1906.01348}}].

\bibitem{Kirkpatrick:2020fwd}
K.~Kirkpatrick, A.~E. Mirasola and C.~Prescod-Weinstein, \emph{{Relaxation
  times for Bose-Einstein condensation in axion miniclusters}},
  \href{https://doi.org/10.1103/PhysRevD.102.103012}{\emph{Phys. Rev. D}
  {\bfseries 102} (2020) 103012}
  [\href{https://arxiv.org/abs/2007.07438}{{\ttfamily 2007.07438}}].

\bibitem{Eby:2020ply}
J.~Eby, L.~Street, P.~Suranyi and L.~C.~R. Wijewardhana, \emph{{Global view of
  axion stars with nearly Planck-scale decay constants}},
  \href{https://doi.org/10.1103/PhysRevD.103.063043}{\emph{Phys. Rev. D}
  {\bfseries 103} (2021) 063043}
  [\href{https://arxiv.org/abs/2011.09087}{{\ttfamily 2011.09087}}].

\bibitem{Kouvaris:2019nzd}
C.~Kouvaris, E.~Papantonopoulos, L.~Street and L.~C.~R. Wijewardhana,
  \emph{{Probing bosonic stars with atomic clocks}},
  \href{https://doi.org/10.1103/PhysRevD.102.063014}{\emph{Phys. Rev. D}
  {\bfseries 102} (2020) 063014}
  [\href{https://arxiv.org/abs/1910.00567}{{\ttfamily 1910.00567}}].

\bibitem{Banerjee:2019epw}
A.~Banerjee, D.~Budker, J.~Eby, H.~Kim and G.~Perez, \emph{{Relaxion Stars and
  their detection via Atomic Physics}},
  \href{https://doi.org/10.1038/s42005-019-0260-3}{\emph{Commun. Phys.}
  {\bfseries 3} (2020) 1} [\href{https://arxiv.org/abs/1902.08212}{{\ttfamily
  1902.08212}}].

\bibitem{Kolb:1993hw}
E.~W. Kolb and I.~I. Tkachev, \emph{{Nonlinear axion dynamics and formation of
  cosmological pseudosolitons}},
  \href{https://doi.org/10.1103/PhysRevD.49.5040}{\emph{Phys. Rev. D}
  {\bfseries 49} (1994) 5040}
  [\href{https://arxiv.org/abs/astro-ph/9311037}{{\ttfamily
  astro-ph/9311037}}].

\bibitem{Eby:2020eas}
J.~Eby, L.~Street, P.~Suranyi, L.~R. Wijewardhana and M.~Leembruggen,
  \emph{{Galactic Condensates composed of Multiple Axion Species}},
  \href{https://arxiv.org/abs/2002.03022}{{\ttfamily 2002.03022}}.

\bibitem{Eby:2018dat}
J.~Eby, M.~Leembruggen, L.~Street, P.~Suranyi and L.~Wijewardhana,
  \emph{{Approximation methods in the study of boson stars}},
  \href{https://doi.org/10.1103/PhysRevD.98.123013}{\emph{Phys. Rev. D}
  {\bfseries 98} (2018) 123013}
  [\href{https://arxiv.org/abs/1809.08598}{{\ttfamily 1809.08598}}].

\bibitem{Pitjev:2013sfa}
N.~Pitjev and E.~Pitjeva, \emph{{Constraints on dark matter in the solar
  system}}, \href{https://doi.org/10.1134/S1063773713020060}{\emph{Astron.
  Lett.} {\bfseries 39} (2013) 141}
  [\href{https://arxiv.org/abs/1306.5534}{{\ttfamily 1306.5534}}].

\bibitem{Adler:2008rq}
S.~L. Adler, \emph{{Placing direct limits on the mass of earth-bound dark
  matter}}, \href{https://doi.org/10.1088/1751-8113/41/41/412002}{\emph{J.
  Phys. A} {\bfseries 41} (2008) 412002}
  [\href{https://arxiv.org/abs/0808.0899}{{\ttfamily 0808.0899}}].

\bibitem{Piazza:2010ye}
F.~Piazza and M.~Pospelov, \emph{{Sub-eV scalar dark matter through the
  super-renormalizable Higgs portal}},
  \href{https://doi.org/10.1103/PhysRevD.82.043533}{\emph{Phys. Rev. D}
  {\bfseries 82} (2010) 043533}
  [\href{https://arxiv.org/abs/1003.2313}{{\ttfamily 1003.2313}}].

\bibitem{Stadnik:2016zkf}
Y.~Stadnik and V.~Flambaum, \emph{{Improved limits on interactions of low-mass
  spin-0 dark matter from atomic clock spectroscopy}},
  \href{https://doi.org/10.1103/PhysRevA.94.022111}{\emph{Phys. Rev. A}
  {\bfseries 94} (2016) 022111}
  [\href{https://arxiv.org/abs/1605.04028}{{\ttfamily 1605.04028}}].

\bibitem{Flacke:2016szy}
T.~Flacke, C.~Frugiuele, E.~Fuchs, R.~S. Gupta and G.~Perez,
  \emph{{Phenomenology of relaxion-Higgs mixing}},
  \href{https://doi.org/10.1007/JHEP06(2017)050}{\emph{JHEP} {\bfseries 06}
  (2017) 050} [\href{https://arxiv.org/abs/1610.02025}{{\ttfamily
  1610.02025}}].

\bibitem{Cosme:2018nly}
C.~Cosme, J.~a.~G. Rosa and O.~Bertolami, \emph{{Scale-invariant scalar field
  dark matter through the Higgs portal}},
  \href{https://doi.org/10.1007/JHEP05(2018)129}{\emph{JHEP} {\bfseries 05}
  (2018) 129} [\href{https://arxiv.org/abs/1802.09434}{{\ttfamily
  1802.09434}}].

\bibitem{Kouvaris:2014uoa}
C.~Kouvaris, I.~M. Shoemaker and K.~Tuominen, \emph{{Self-Interacting Dark
  Matter through the Higgs Portal}},
  \href{https://doi.org/10.1103/PhysRevD.91.043519}{\emph{Phys. Rev. D}
  {\bfseries 91} (2015) 043519}
  [\href{https://arxiv.org/abs/1411.3730}{{\ttfamily 1411.3730}}].

\bibitem{Sirunyan:2018owy}
{\scshape CMS} collaboration, \emph{{Search for invisible decays of a Higgs
  boson produced through vector boson fusion in proton-proton collisions at
  $\sqrt{s} =$ 13 TeV}},
  \href{https://doi.org/10.1016/j.physletb.2019.04.025}{\emph{Phys. Lett. B}
  {\bfseries 793} (2019) 520}
  [\href{https://arxiv.org/abs/1809.05937}{{\ttfamily 1809.05937}}].

\bibitem{Scherrer:1992na}
R.~J. Scherrer and D.~N. Spergel, \emph{{How constant is the Fermi coupling
  constant?}}, \href{https://doi.org/10.1103/PhysRevD.47.4774}{\emph{Phys. Rev.
  D} {\bfseries 47} (1993) 4774}.

\bibitem{Schlamminger:2007ht}
S.~Schlamminger, K.-Y. Choi, T.~Wagner, J.~Gundlach and E.~Adelberger,
  \emph{{Test of the equivalence principle using a rotating torsion balance}},
  \href{https://doi.org/10.1103/PhysRevLett.100.041101}{\emph{Phys. Rev. Lett.}
  {\bfseries 100} (2008) 041101}
  [\href{https://arxiv.org/abs/0712.0607}{{\ttfamily 0712.0607}}].

\bibitem{Bertotti:2003rm}
B.~Bertotti, L.~Iess and P.~Tortora, \emph{{A test of general relativity using
  radio links with the Cassini spacecraft}},
  \href{https://doi.org/10.1038/nature01997}{\emph{Nature} {\bfseries 425}
  (2003) 374}.

\bibitem{Kapner:2006si}
D.~Kapner, T.~Cook, E.~Adelberger, J.~Gundlach, B.~R. Heckel, C.~Hoyle et~al.,
  \emph{{Tests of the gravitational inverse-square law below the dark-energy
  length scale}},
  \href{https://doi.org/10.1103/PhysRevLett.98.021101}{\emph{Phys. Rev. Lett.}
  {\bfseries 98} (2007) 021101}
  [\href{https://arxiv.org/abs/hep-ph/0611184}{{\ttfamily hep-ph/0611184}}].

\bibitem{PhysRevLett.123.091601}
B.~Grinstein, C.~Kouvaris and N.~G. Nielsen, \emph{Neutron star stability in
  light of the neutron decay anomaly},
  \href{https://doi.org/10.1103/PhysRevLett.123.091601}{\emph{Phys. Rev. Lett.}
  {\bfseries 123} (2019) 091601}.

\bibitem{JUNGMAN1996195}
G.~Jungman, M.~Kamionkowski and K.~Griest, \emph{Supersymmetric dark matter},
  \href{https://doi.org/https://doi.org/10.1016/0370-1573(95)00058-5}{\emph{Physics
  Reports} {\bfseries 267} (1996) 195}.

\bibitem{HAIYANGCHENG1989347}
{Hai-Yang Cheng}, \emph{Low-energy interactions of scalar and pseudoscalar
  higgs bosons with baryons},
  \href{https://doi.org/https://doi.org/10.1016/0370-2693(89)90402-4}{\emph{Physics
  Letters B} {\bfseries 219} (1989) 347}.

\bibitem{GASSER1991252}
J.~Gasser, H.~Leutwyler and M.~Sainio, \emph{Sigma-term update},
  \href{https://doi.org/https://doi.org/10.1016/0370-2693(91)91393-A}{\emph{Physics
  Letters B} {\bfseries 253} (1991) 252}.

\bibitem{Primack:1973iz}
J.~R. Primack and H.~R. Quinn, \emph{{Muon g-2 and other constraints on a model
  of weak and electromagnetic interactions without neutral currents}},
  \href{https://doi.org/10.1103/PhysRevD.6.3171}{\emph{Phys. Rev. D} {\bfseries
  6} (1972) 3171}.

\bibitem{Bardeen:1972vi}
W.~A. Bardeen, R.~Gastmans and B.~E. Lautrup, \emph{{Static quantities in
  Weinberg's model of weak and electromagnetic interactions}},
  \href{https://doi.org/10.1016/0550-3213(72)90218-0}{\emph{Nucl. Phys. B}
  {\bfseries 46} (1972) 319}.

\bibitem{Leveille:1977rc}
J.~P. Leveille, \emph{{The Second Order Weak Correction to (G-2) of the Muon in
  Arbitrary Gauge Models}},
  \href{https://doi.org/10.1016/0550-3213(78)90051-2}{\emph{Nucl. Phys. B}
  {\bfseries 137} (1978) 63}.

\bibitem{Haber:1978jt}
H.~E. Haber, G.~L. Kane and T.~Sterling, \emph{{The Fermion Mass Scale and
  Possible Effects of Higgs Bosons on Experimental Observables}},
  \href{https://doi.org/10.1016/0550-3213(79)90225-6}{\emph{Nucl. Phys. B}
  {\bfseries 161} (1979) 493}.

\bibitem{Carlson:1988dp}
E.~D. Carlson, S.~L. Glashow and U.~Sarid, \emph{{Searching for a Light
  Higgs}}, \href{https://doi.org/10.1016/0550-3213(88)90331-8}{\emph{Nucl.
  Phys. B} {\bfseries 309} (1988) 597}.

\bibitem{Zhou:2001ew}
Y.-F. Zhou and Y.-L. Wu, \emph{{Lepton flavor changing scalar interactions and
  muon g-2}}, \href{https://doi.org/10.1140/epjc/s2003-01137-1}{\emph{Eur.
  Phys. J. C} {\bfseries 27} (2003) 577}
  [\href{https://arxiv.org/abs/hep-ph/0110302}{{\ttfamily hep-ph/0110302}}].

\bibitem{Stadnik:2015kia}
Y.~Stadnik and V.~Flambaum, \emph{{Can dark matter induce cosmological
  evolution of the fundamental constants of Nature?}},
  \href{https://doi.org/10.1103/PhysRevLett.115.201301}{\emph{Phys. Rev. Lett.}
  {\bfseries 115} (2015) 201301}
  [\href{https://arxiv.org/abs/1503.08540}{{\ttfamily 1503.08540}}].

\bibitem{Stadnik:2015uka}
Y.~Stadnik and V.~Flambaum, \emph{{Constraining scalar dark matter with Big
  Bang nucleosynthesis and atomic spectroscopy}},
  \href{https://arxiv.org/abs/1504.01798}{{\ttfamily 1504.01798}}.

\bibitem{Weyers_2018}
S.~Weyers, V.~Gerginov, M.~Kazda, J.~Rahm, B.~Lipphardt, G.~Dobrev et~al.,
  \emph{Advances in the accuracy, stability, and reliability of the {PTB}
  primary fountain clocks},
  \href{https://doi.org/10.1088/1681-7575/aae008}{\emph{Metrologia} {\bfseries
  55} (2018) 789}.

\bibitem{Han_2019}
J.~Han, Y.~Zuo, J.~Zhang and L.~Wang, \emph{Theoretical investigation of the
  black-body zeeman shift for microwave atomic clocks},
  \href{https://doi.org/10.1140/epjd/e2018-90342-1}{\emph{The European Physical
  Journal D} {\bfseries 73} (2019) }.

\bibitem{PhysRevLett.104.070802}
C.~W. Chou, D.~B. Hume, J.~C.~J. Koelemeij, D.~J. Wineland and T.~Rosenband,
  \emph{Frequency comparison of two high-accuracy ${\mathrm{al}}^{+}$ optical
  clocks}, \href{https://doi.org/10.1103/PhysRevLett.104.070802}{\emph{Phys.
  Rev. Lett.} {\bfseries 104} (2010) 070802}.

\bibitem{Bloom:2013uoa}
B.~J. Bloom, T.~L. Nicholson, J.~R. Williams, S.~L. Campbell, M.~Bishof,
  X.~Zhang et~al., \emph{{An Optical Lattice Clock with Accuracy and Stability
  at the $10^{-18}$ Level}},
  \href{https://doi.org/10.1038/nature12941}{\emph{Nature} {\bfseries 506}
  (2014) 71} [\href{https://arxiv.org/abs/1309.1137}{{\ttfamily 1309.1137}}].

\bibitem{PhysRevLett.120.103201}
G.~E. Marti, R.~B. Hutson, A.~Goban, S.~L. Campbell, N.~Poli and J.~Ye,
  \emph{Imaging optical frequencies with $100\text{ }\text{
  }\ensuremath{\mu}\mathrm{Hz}$ precision and $1.1\text{ }\text{
  }\ensuremath{\mu}\mathrm{m}$ resolution},
  \href{https://doi.org/10.1103/PhysRevLett.120.103201}{\emph{Phys. Rev. Lett.}
  {\bfseries 120} (2018) 103201}.

\bibitem{PhysRevLett.108.120802}
C.~J. Campbell, A.~G. Radnaev, A.~Kuzmich, V.~A. Dzuba, V.~V. Flambaum and
  A.~Derevianko, \emph{Single-ion nuclear clock for metrology at the 19th
  decimal place},
  \href{https://doi.org/10.1103/PhysRevLett.108.120802}{\emph{Phys. Rev. Lett.}
  {\bfseries 108} (2012) 120802}.

\bibitem{Peik:2020cwm}
E.~Peik, T.~Schumm, M.~S. Safronova, A.~P\'alffy, J.~Weitenberg and P.~G.
  Thirolf, \emph{{Nuclear clocks for testing fundamental physics}},
  \href{https://doi.org/10.1088/2058-9565/abe9c2}{\emph{Quantum Sci. Technol.}
  {\bfseries 6} (2021) 034002}
  [\href{https://arxiv.org/abs/2012.09304}{{\ttfamily 2012.09304}}].

\bibitem{McCabe:2013kea}
C.~McCabe, \emph{{The Earth's velocity for direct detection experiments}},
  \href{https://doi.org/10.1088/1475-7516/2014/02/027}{\emph{JCAP} {\bfseries
  02} (2014) 027} [\href{https://arxiv.org/abs/1312.1355}{{\ttfamily
  1312.1355}}].

\bibitem{Emken:2017qmp}
T.~Emken and C.~Kouvaris, \emph{{DaMaSCUS: The Impact of Underground
  Scatterings on Direct Detection of Light Dark Matter}},
  \href{https://doi.org/10.1088/1475-7516/2017/10/031}{\emph{JCAP} {\bfseries
  10} (2017) 031} [\href{https://arxiv.org/abs/1706.02249}{{\ttfamily
  1706.02249}}].

\bibitem{Celestrak}
T.~Kelso, \emph{Celestrak (available at https://celestrak.com/)}, .

\end{thebibliography}\endgroup
\bibliographystyle{JHEP}

\end{document}